\documentclass[reprint,aps,prl,superscriptaddress,showpacs,longbibliography]{revtex4-1}

\usepackage[pdftex]{graphicx} \usepackage{float}

\usepackage{amsmath} \usepackage{amsfonts} \usepackage{amssymb}
\usepackage{wasysym} \usepackage[pdftex,colorlinks=true]{hyperref}
\usepackage{mathtools} \usepackage{color}

\DeclarePairedDelimiterX\expval[1]{\langle}{\rangle}{#1}

\usepackage[pdftex,dvipsnames]{xcolor}
\usepackage[colorinlistoftodos,prependcaption,textsize=tiny]{todonotes}
\definecolor{orange}{rgb}{1,0.5,0}

\newcommand{\vect}[1]{\mathbf{#1}}

\def\sectionn#1{\noindent\underline{\it #1:}}

\AtBeginDocument{%
  \newwrite\bibnotes \def\bibnotesext{Notes.bib}
  \immediate\openout\bibnotes=\jobname\bibnotesext
  \immediate\write\bibnotes{@CONTROL{REVTEX41Control}}
  \immediate\write\bibnotes{@CONTROL{%
      apsrev41Control,author="08",editor="1",pages="1",title="0",year="1"}}
  \if@filesw \immediate\write\@auxout{\string\citation{apsrev41Control}}%
  \fi }%

\begin{document}

\title{Temperature dependence of butterfly effect in a classical many-body
  system}
% \title{Butterfly effect in the classical $\mathbb{Z}_2$ spin liquid}

\author{Thomas Bilitewski} \affiliation{Max-Planck-Institut f\"{u}r Physik
  komplexer Systeme, N\"othnitzer Str.\ 38, 01187 Dresden, Germany}

\author{Subhro Bhattacharjee} \affiliation{International Centre for Theoretical
  Sciences, Tata Institute of Fundamental Research, Bengaluru 560089, India}

\author{Roderich Moessner} \affiliation{Max-Planck-Institut f\"{u}r Physik
  komplexer Systeme, N\"othnitzer Str.\ 38, 01187 Dresden, Germany}

\begin{abstract}
  We study the chaotic dynamics in a classical many-body system of interacting
  spins on the kagome lattice. We characterise many-body chaos via the butterfly
  effect as captured by an appropriate out-of-time-ordered correlator. Due to
  the emergence of a spin liquid phase, the chaotic dynamics extends all the way
  to zero temperature. We thus determine the full temperature dependence of two
  complementary aspects of the butterfly effect: the Lyapunov exponent, $\mu$,
  and the butterfly speed, $v_b$, and study their interrelations with usual
  measures of spin dynamics such as the spin-diffusion constant, $D$ and
  spin-autocorrelation time, $\tau$. We find that they all exhibit power law
  behaviour at low temperature, consistent with scaling of the form $D\sim
  v_b^2/\mu$ and $\tau^{-1}\sim T$. The vanishing of $\mu\sim T^{0.48}$ is
  parametrically slower than that of the corresponding quantum bound, $\mu\sim
  T$, raising interesting questions regarding the semi-classical limit of such
  spin systems.
\end{abstract}

\maketitle
%%%%%%%%%%%%%%%%%%%%%%%%
\sectionn{Introduction} %
Chaos
\cite{Dellago1997,Hoover2002,Latora1998,Zon1998,Beijeren1997,Wijn2004,Wijn2005,Wijn2012,Wijn2013,gaspard1998experimental,dettmann1999statistical}
underpins much of statistical mechanics, providing the basis for ergodicity,
thermalization and transport in many-body systems. Perhaps its most striking
feature that has captured public imagination is the butterfly effect
\cite{lorenz1963deterministic, lorenz1996essence, lorenz2000butterfly,
  hilborn2004sea}: an infinitesimal local change of initial condition is
amplified exponentially (Lyapunov exponent $\mu$) and spreads out ballistically
(butterfly speed $v_b$) to dramatically affect global outcomes.

Quantitative connections between characteristic time and length scales of the
chaotic dynamics of a many-body system and those related to its thermalization
and transport are far from settled, receiving renewed attention particularly for
quantum many-body systems
\cite{Maldacena2016,Hosur2016,Roberts2015,Roberts2015a,Shenker2014,Cotler2017,Dora2017,Menezes2018,Rozenbaum2017,Luitz2017,Kukuljan2017,Aleiner2016,Lin2018,Chen2016,Chan2017,Nahum2018,Keyserlingk2018,Rakovszky2017,Khemani2017,Stanford2016,Patel2017,Chowdhury2017}.
There, diagnostic tools of chaos akin to $\mu$ and $v_b$ were obtained in an
appropriately defined limit of {\it out-of-time-ordered commutators} (OTOC)
\cite{Srednicki1994,Gutzwiller1991,Larkin1969}. Further, in a recent study of
classical spin chain at infinite temperature \cite{Das2018}, the classical limit
of OTOC's has been shown to characterize these features of the butterfly effect.

Here, we study the evolution of the chaotic dynamics as a function of
temperature, $T$, of the many body system, and its interrelation with
thermalization and transport quantities such as relaxation and diffusion. This
is interesting as correlations develop due to interactions as $T$ is lowered,
thereby affecting the dynamics of the system. Generally, one expects that at low
$T$, the effect of chaos may be weakened due to emergence of long-lived
quasi-particles (with or without spontaneously broken symmetry) that dominate
the dynamics. Indeed, recent studies of the quantum Sachdev-Ye-Kitaev (SYK)
\cite{PhysRevLett.70.3339, kitaev,PhysRevB.95.134302,PhysRevLett.119.216601} and
finite density fermions coupled to gauge fields \cite{Patel1844} (both in the
large $N$ limit), show that the absence of quasi-particles due to interactions
can lead to chaos as manifested in OTOCs even at the lowest $T$.

We explore these issues in an interacting many-body classical spin system with
local interactions on a two dimensional kagome lattice. We elucidate the
$T$-dependence of Lyapunov exponent and butterfly speed and find connections
between diffusion and chaos over the entire temperature range. At low $T$,
Lyapunov exponent ($\mu \sim T^{0.48}$) and butterfly speed ($v_b \sim
T^{0.23}$), extracted from a classical OTOC \cite{Das2018}, show novel algebraic
scaling with $T$. This behaviour is qualitatively distinct from that of the
quantum counterparts since the observed sub-linear scaling of the Lyapunov
exponent is at odds with the {\it quantum low-$T$ bound} ($\mu \le 2 \pi k_B
T/\hbar$ \cite{Maldacena2016}). This raises questions regarding the presumably
singular nature of semi-classical (in $1/S$ sense) corrections
\cite{Kurchan2016,Scaffidi2017} in this spin system. However, in spite of the
seeming ``violation'' of the quantum bound, our results are consistent with a
recently identified connections of these microscopic measures of chaos and the
macroscopic phenomenon of transport, where for the SYK-model and ``strange''
metals the energy diffusion constant was found to scale as $D \sim
v_b^2/\mu$\cite{Gu2017,Blake2016,Blake2017,Werman2017,Lucas2017}.
\begin{figure}
  \begin{minipage}{0.49\columnwidth}
    \includegraphics[width=.99\columnwidth]{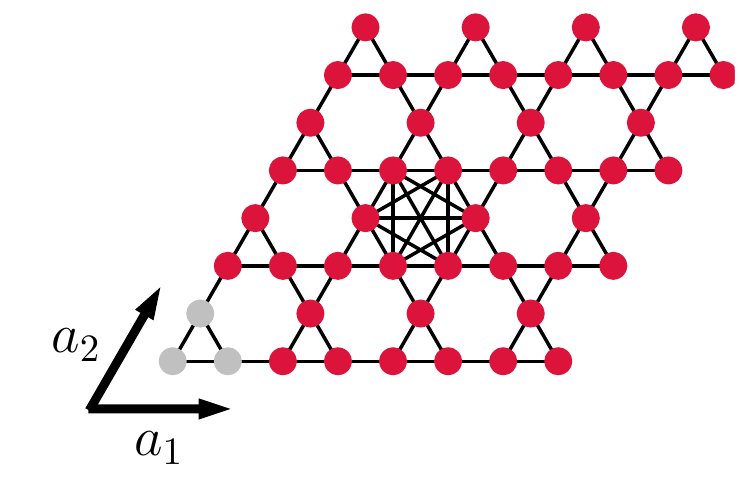}
  \end{minipage}
  \begin{minipage}{0.49\columnwidth}
    \includegraphics[width=.99\columnwidth]{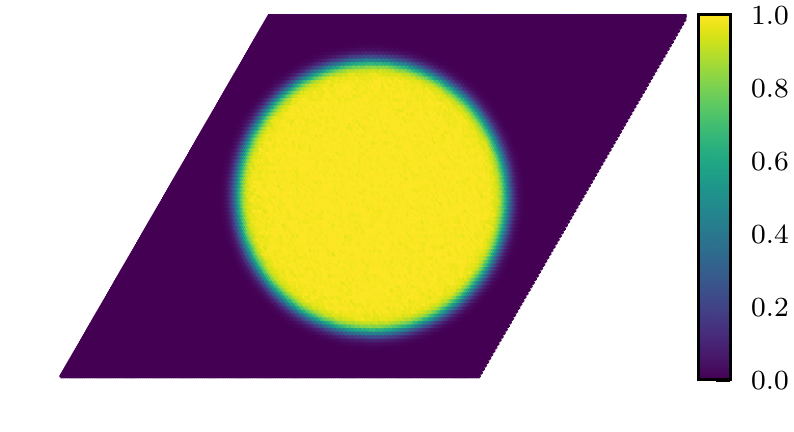}
  \end{minipage}
  \caption{Left: The classical Heisenberg model hosting the $\mathbb{Z}_2$ spin liquid is defined on the Kagome lattice with
    couplings fully connecting all hexagons (only shown for central one).
    Indicated are the basis vectors $a_1,a_2$ and a possible
    unit cell (light gray vertices).
   Right: Snapshot of the dynamics of the de-correlator $D(x,t)$ showing
   ballistic isotropic spreading of a perturbation initially localised in the
   centre of the system.
   \label{fig:illustration_lattice}}
\end{figure}

Persistence of chaotic dynamics, usually characteristic to high $T$, in our spin
system all the way down to $T=0$ owes its origin to competing (frustrated) local
interactions that completely suppresses magnetic ordering.%
%%%%%%%%%%%%%%%%

\sectionn{Model} We study classical $O(3)$ Heisenberg spins of unit length,
${\bf S}_{\bf x}$, on the sites ${\bf x}$ of the kagome lattice,
\begin{equation} H= J\sum_{ {\bf x,x'}\in\hexagon} \vect{S}_{\bf x}\cdot
  \vect{S}_{\bf x'} = \frac{J}{2} \sum_{\alpha} (\mathbf{L}_{\alpha})^2 +
  \textrm{const} ,
  \label{eq:H_Z2}
\end{equation}
where each spin interacts equally with all the spins with which it shares a
hexagon, $\hexagon$ \cite{PhysRevB.65.224412}, whose total spin is denoted by
$\mathbf{L}_{\alpha} = \sum_{\bf{x} \in \hexagon}
\mathbf{S}_{\bf{x}}$, % is the total spin sum on a hexagon $\alpha$,
schematically illustrated in Fig.~\ref{fig:illustration_lattice}.

For antiferromagnetic interactions ($J>0$), ground states satisfy the local
constraints $\mathbf{L}_{\alpha}=0$ for each hexagon, which leads to a
macroscopically degenerate ground state manifold. The system remains in a
paramagnetic state all the way down to $T=0$ which has a finite spin correlation
length and exhibits fractionalization \cite{Rehn2017}. Interestingly, the system
does not freeze or fall out of equilibrium in the entire temperature range. Such
a phase has been dubbed a classical `$\mathbb{Z}_2$' spin-liquid.

% with orphan fractionalisation, exponentially decaying correlations down to
% $T=0$, and no order-by-disorder . {\color{blue}{\bf [comment about the
% structure of the ground state manifold ?]}}}

The dynamics is that of spins precessing around their local exchange fields,
which conserves total energy $E$, magnetization $M$, as well as the spin norm:
\begin{equation} \frac{d \vect{S}_{\bf{x}}(t)}{dt} = - \vect{S}_{\bf{x}}(t)
  \times \sum_{j} J_{\bf{x} \bf{x'}} \vect{S}_{\bf{x'}}(t) \, .
  \label{eq:bloch_eq}
\end{equation}
%%%%%%%%%%%%%%%%%%%%%%%%%%%%%%%

\sectionn{Numerical simulations} %
These were performed over a range of temperature $T=10^{-3}$ to $T=100$, and
linear system size $L=25$ to $L=201$ with $N_s=3 L^2$ spins and periodic
boundary conditions. Results shown are for $L=101$ unless indicated otherwise.
The spin dynamics is integrated using an eighth-order Runge-Kutta solver with a
time-step chosen such that energy/site and magnetisation/site are conserved to
better than $\sim 10^{-8}$. Results are averaged over $10^4$ initial states
sampled from the Boltzmann distribution via Monte-Carlo.
% The MC-simulations use heat-bath and microcanonical over-relaxation updates
% and the number of MC-steps between measurements is dynamically adjusted to
% ensure an overlap $<0.01$ between successive spin configurations. Results for
% the de-correlator $D(x,t)$ are defined along 1D-cuts starting at the perturbed
% site and averaged over the symmetry-related directions.
Details on the fitting procedure and exemplary raw data fits can be found in the
suppl.mat. \cite{supplemental}. We measure energy in units of $J=1$, and
distances in units of the lattice spacing $a=1$.
%%%%%%%%%

\sectionn{Temperature dependence of dynamics} %
We begin by discussing the two point spin correlator
\begin{align}
  C({\bf x},t)=\langle {\bf S_x}(t)\cdot {\bf S_0}(0)\rangle \,,
\end{align}
its Fourier-transform, the dynamical structure factor
$\mathcal{S}(\vect{q},\omega)$, and the auto-correlator $A(t) = \sum_{\bf{x}}
\langle {\bf S_x}(t)\cdot {\bf S_x}(0)\rangle$.

\begin{figure}
  \begin{minipage}{0.99\columnwidth}
    \includegraphics[width=.99\columnwidth]{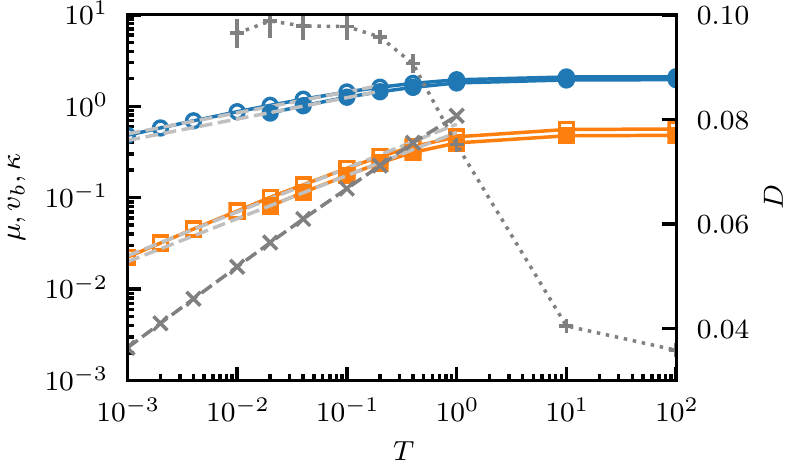}
  \end{minipage}
  \caption{Temperature dependence of various quantities characterising the
    dynamics and chaos.
     Relaxation rate $\kappa$ (`x') of the auto-correlation function $A(t)\sim
    e^{-\kappa t}$ on a log-scale (left y-axis) and diffusion constant $D$
    ('+') on a linear scale (right y-axis).
     Lyapunov exponent $\mu$ (squares) and butterfly velocity
    $v_b$ (circles).
    Filled symbols results from fits to the scaling form,
     Eq.~\ref{eq:D_scaling}, empty symbols obtained from independent fits to $D(x=0,t)$ for $\mu$, see Fig.
     \ref{fig:D_scaling_form}, and the arrival times $t_{D_0}$ for $v_b$, Gray dashed lines are the power-laws fitted to the
     low-temperature regime. We extract powerlaws $\mu \sim T^{0.48}$ (``ind''),
     $\mu \sim T^{0.47}$ (``fit'') and $v_b \sim T^{0.23}$ in
     the low-temperature regime respectively.
     \label{fig:temperature_dependence}}
 \end{figure}

 At all $T$, the autocorrelator exhibits an initial exponential decay $A(t)\sim
 e^{-\kappa t}$, with diffusion at long wavelengths seen in the tail at long
 times, $A(t) \sim 1/t$, as well as in the decay of the dynamical structure
 factor close to the $\Gamma$-point, $\mathcal{S}(\vect{q},\omega) \sim 1/[(D
 q^2)^2+\omega^2]$ \cite{supplemental}.

 The $T$ dependence of relaxation rate $\kappa$ and diffusion constant $D$ is
 shown Fig.~\ref{fig:temperature_dependence}. We observe a linear scaling
 $\kappa \sim T$, and saturation of the diffusion constant to a constant value,
 in the low $T$ spin-liquid regime in conformity with the large-$N$
   results \cite{supplemental}.

 Having established and characterised the diffusive behaviour of the spin
 correlators, we now turn to the main subject of this work, the many-body chaos
 in this many-body system, in the form of the butterfly effect.

 \sectionn{OTOC} %
 We characterise chaos using an analogue of the OTOC in classical spin systems
 that was constructed in Ref.\cite{Das2018}. Considering the evolution of two
 copies with slightly perturbed initial conditions, we define
 \begin{equation} D(x,t) = 1 - \expval{\vect{S}_x(t)\cdot \vect{\tilde{S}}_x(t)}
   = \expval{\left[ \delta \vect{S}_x(t)\right]^2}/2
 \end{equation}
 with cross-correlator $\expval{\vect{S} \cdot \vect{\tilde{S}}}$ between
 copies, the perturbed spin configuration $\vect{\tilde{S}} = \vect{S} + \delta
 \vect{S}$, and $\expval{\cdots}$ an average over the thermodynamic ensemble at
 $T$.

 % We will also make use of a linearised version of the dynamical equations,
 % defined for the difference of the perturbed and original spin configuration,
 % $\delta \vect{S}_i = \vect{\tilde{S}}_i - \vect{S}_i$, which approximately
 % evolves as
 % \begin{equation}
 %   \frac{\delta \vect{S}_i(t)}{dt} = - \delta \vect{S}_i(t) \times
 %   \sum_{j}J_{ij} \vect{S}_j(t) - \vect{S}_i(t) \times \sum_{j}J_{ij}
 %   \delta\vect{S}_j(t) \, ,
 % \end{equation}
 % which is valid in the limit of $\epsilon \to 0$.

 This de-correlator $D(x,t)$ is expected to scale as
 \begin{equation}
   D(x,t) \sim \exp [2 \mu (1- (v/v_b)^{\nu}) t]
   \label{eq:D_scaling}
 \end{equation}
 with Lyapunov exponent $\mu$, butterfly velocity $v_b$ and an exponent $\nu$,
 in general all $T$-dependent. The exponent $\nu$ defines the functional form of
 the velocity-dependent Lyapunov exponent $\lambda(v) = \mu (1-(v/v_b)^{\nu})$
 \cite{Kaneko1986,Deissler1984,Deissler1987,Das2018,Khemani2018,Xu2018}, which
 measures the exponential growth rate of the de-correlator along rays $v=x/t$.
 It depends both on the dimensionality, typically decreasing in larger
 dimensions with decreasing (quantum) fluctuations, and on the presence and
 type/range of interactions
 \cite{Xu2018,Khemani2018,Nahum2018,Keyserlingk2018,Lin2018a}.

 For full quantum models, the Lyapunov exponent inside the light-cone ($v<v_b$)
 has been found to be zero in several examples, whereas in large-N and
 (semi-)classical models a regime of exponential growth is possible
 \cite{Khemani2018}.
%%%%%%%%%%%%%%%%%%%%%%%%%%%%%%%%

 % The results from the non-linear dynamics is compatible with an exponent
 % $\nu=2$ over the full $T$ range, whereas the linearised dynamics shows a
 % slight decrease from $\nu=2$ at $T=100$ to about $\nu=1.9$ at $T=0.1$.
 % However, the error on the exponent extracted from the non-linear data is
 % rather large, since the data only approximately collapses in a region $v \sim
 % v_b$, and the data can equally well be fit with the exponent obtained from
 % the linearised dynamics.

 \sectionn{Ballistic Spread of Decorrelation} %
 A snapshot of the decorrelation wavefront at a particular time, as measured by
 OTOC, is shown in Fig.~\ref{fig:illustration_lattice}. A light-cone is visible
 in the dynamics throughout the entire temperature range separating a
 decorrelated region with $D \sim 1$ centered at the initially perturbed site
 from a fully correlated unperturbed region with $D \sim 0$. The speed of the
 ballistally propagating light-cone of the perturbation allows us to define the
 butterfly speed, $v_b$ \cite{Lieb1972,Marchioro1978,Metivier2014,Roberts2016}.
 The wave-front after an initial transient remains circular over the course of
 the dynamics and the full temperature range we consider eventhough the
 underlying lattice only has a six-fold rotational symmetry. This constitutes a
 non-trivial model-dependent feature of OTOC's, which generically only need to
 respect, even at late times, the discrete lattice symmetries \cite{Nahum2018}.
 % However, we cannot exclude that on larger systems and for longer times, a
 % different asymptotic shape could emerge once potentially present small
 % symmetry-breaking terms become significant.
 Thus, it is sufficient to restrict to 1D cuts in the following discussions.

 \sectionn{Temperature dependence of $v_b$ and $\mu$} %
 The next central result of this study, the full $T$ dependence of Lyapunov
 exponent $\mu$ and the butterfly velocity $v_b$, is shown in
 Fig.~\ref{fig:temperature_dependence}. The exponential growth of the
 decorrelation throughout the entire temperature range confirms the persistence
 of the chaotic dynamics down to the lowest $T$, in keeping with the
 persistence of the spin liquid phase.

 We employ two methods to extract $\mu$ and $v_b$, a fit to the scaling form of
 the wavefronts, Eq.~\ref{eq:D_scaling}, and independent fits discussed in
 detail below. Generically, the Lyapunov exponent and butterfly speed extracted
 from the fit to the full scaling form are slightly lower than those determined
 independently, which we attribute to subleading prefactors not contained in the
 scaling form (Eq.~\ref{eq:D_scaling}). Additionally, the independent fits can
 be extended to lower temperatures than fitting the full wavefront.
 
 For both methods we observe algebraic behaviour at low temperatures: the
 butterfly speed scales as $\mu\sim T^{0.48\pm 0.006}$ ($T^{0.47\pm 0.005}$) for
 the independent fit (scaling form) and the Lyapunov exponent as $v_b \sim
 T^{0.23\pm 0.01}$.

 The observed scaling of the Lyapunov exponent, $\mu \sim T^{0.48}$, is
 parametrically larger at low temperatures than the bound on quantum chaos $\mu
 \le k_B T/h$ \cite{Maldacena2016}. While this bound is not directly applicable
 to our classical model, it implies the semi-classical (say in the form of 1/S)
 corrections are singular in the low $T$ limit \cite{Kurchan2016,Scaffidi2017}.
 At the same time, we note that the observed scaling is consistent with a
 recently suggested $\sqrt{E}$ behaviour \cite{Kurchan2016}.
 
 Interestingly, however, $v_b^2/\mu$ is approximately constant in the low $T$
 regime. This is consistent with the conjectured relation between the diffusion
 constant $D$, the Lyapunov exponent and butterfly velocity as $D \sim
 v_b^2/\mu$ \cite{Gu2017,Blake2016,Blake2017,Werman2017,Lucas2017}, and the fact
 that we obtain a $T$-independent diffusion constant in the spin-liquid regime
 (Fig.~\ref{fig:temperature_dependence}).
 
 We now turn to a detailed discussion of the wavefronts, their ballistic
 propagation, and scaling form. This also illustrates how the discussed
 quantities are obtained from the OTOC.

 \sectionn{Scaling form} %
 \begin{figure}
   \begin{minipage}{0.99\columnwidth}
     \includegraphics[width=.99\columnwidth]{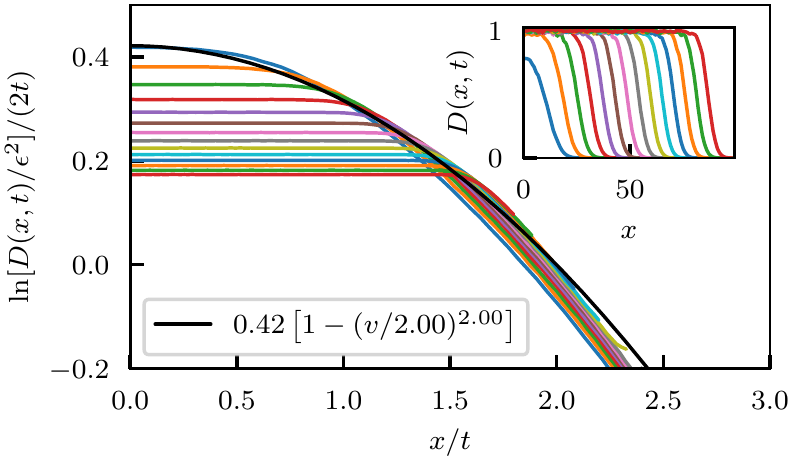}
   \end{minipage}
   \begin{minipage}{0.99\columnwidth}
     \includegraphics[width=.99\columnwidth]{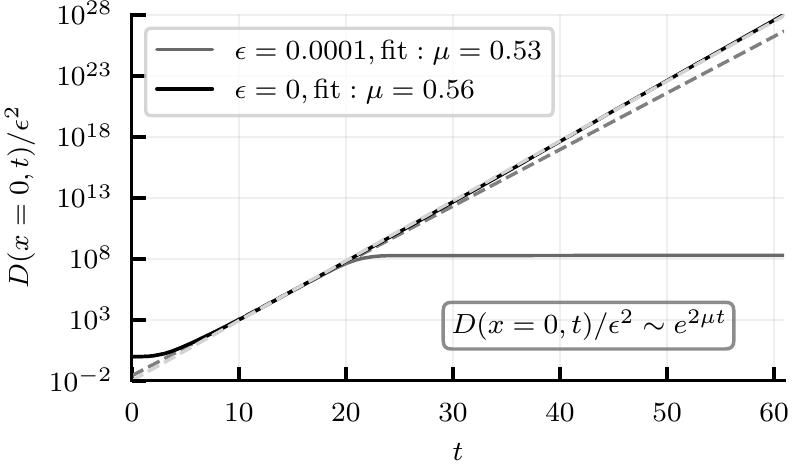}
   \end{minipage}
  \caption{%
    Top: Scaling form and ballistic propagation of the wave-front in the de-correlator $D(x,t)$.  The 
    data at the wave-front $v =x/t \sim v_b$ well fits Eq.~\ref{eq:D_scaling}
    with the parameters given in the box. The inset shows the unscaled data
    versus distance $x$ at different time-slices $t$ demonstrating the ballistic
    propagation of the wave-front after an initial growth period. 
    Bottom: De-correlator $D(x=0,t)$ at the initial perturbed site versus time
    showing exponential growth.
    Data obtained at $T=100$ on a $L=101$ system, for the non-linear dynamics
    with $\epsilon=10^{-4}$ and averaged over $10^4$ initial states.
    \label{fig:D_scaling_form}}
\end{figure}
Fig.~\ref{fig:D_scaling_form} (top panel) shows the scaling form of the
de-correlator, according to Eq.~\ref{eq:D_scaling}, and the build-up and
propagation of the wavefronts (inset). Close to the wavefront we observe
approximate scaling collapse of the de-correlator. In contrast, inside the
light-cone we observe deviations from the scaling form due to the saturation of
the bounded de-correlators. This is avoided in the linearised version of the
dynamics \cite{Das2018}, which allows us to directly access the limit of
vanishingly small perturbation strength $\epsilon$.

\sectionn{Individual fits} Complementary to fitting the full spatiotemporal
profile of $D(x,t)$, which allows access to the velocity dependent Lyapunov
exponent $\lambda(v)$, we extract $\mu = \lambda(v=0)$ from an exponential fit
to $D(x=0,t)$ and the butterfly velocity $v_b$ from the arrival times
$t_{D_0}(x)$ of the wave-front via $v_b=x/t_{D_0}$, where $D(x,t_{D_0})>D_0$
\cite{Das2018}.

We demonstrate the expected exponential growth $D(x=0,t) \sim e^{2 \mu t}$ in
the lower panel of Fig.~\ref{fig:D_scaling_form} comparing the linearised
dynamics to the non-linear dynamics with $\epsilon=10^{-4}$. In particular, the
linearised dynamics shows exponential growth for all times, whereas the
linearised dynamics only shows exponential growth over a finite time increasing
with smaller perturbation strength as $\log(\epsilon)$ before the decorrelator
saturates.

The results for $v_b$ and $\mu$ from the non-linear dynamics converge with
decreasing perturbation strength $\epsilon$ to the results obtained from the
linearised dynamics \cite{supplemental}.

We find that the results obtained from the individual fits are generally
compatible with the results obtained from fitting to the scaling form.
Importantly, it does allow us to reach lower temperatures, where a full
wavefront cannot develop on available system sizes before the perturbation
reaches the periodic boundaries.

\sectionn{Solitonic wavefronts} %
\begin{figure}
  \begin{minipage}{0.99\columnwidth}
    \includegraphics[width=.99\columnwidth]{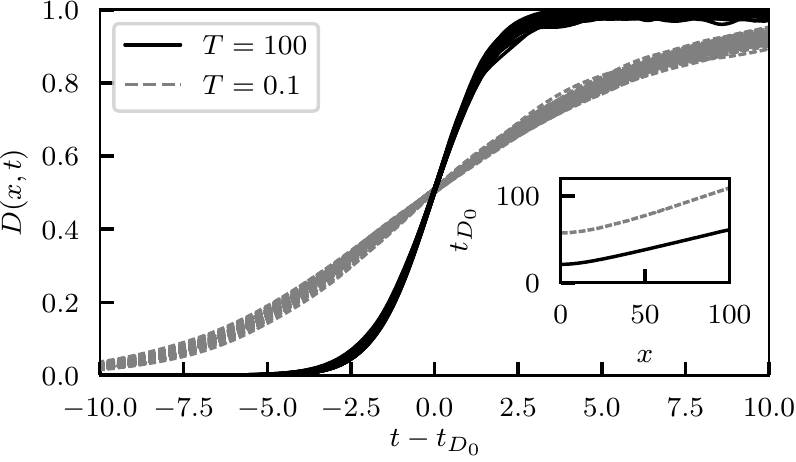}
  \end{minipage}
  \caption{Shape of the propagating wave-front in the de-correlator
     $D(x,t)$ for temperatures $T=100$ (solid black) and $T=0.1$ (dashed gray). Data at
     different distances $x$ for $x=10,\cdots,L-10$ are plotted versus time $t$ after
     collapsing them by subtracting $t_{D_0}(x)$ with $D_0 = 0.5$. Inset:
     Arrival time $t_{D_0}$ versus distance $x$ showing linear scaling with distance.
     \label{fig:wavefronts}}
 \end{figure}
 Fig.~\ref{fig:wavefronts} demonstrates that, like in a soliton, the wavefront
 shape remains approximately constant as it moves. At least for the times
 accessible in our simulation, we do not observe significant broadening. Note
 that here we consider the shape of the full wavefront, rather than the leading
 edge only, which in principle could show different behaviour due to the
 non-linearity of the dynamics.

 In contrast, we observe strong $T$ dependence of the shape of the wavefront.
 With decreasing temperatures the front broadens both temporally,
 Fig.~\ref{fig:wavefronts}, and spatially.
 % (\cite{supplemental}).
 
 The inset of Fig.~\ref{fig:wavefronts} shows the linear scaling of the arrival
 times $t_{D_0}(x)$, with distance $x$, and, thus, ballistic propagation of the
 wavefront. The slowing down of propagation at lower $T$ manifests itself in the
 larger slope of the arrival times versus distance, whereas the decrease of the
 Lyapunov exponent with temperature is seen in the larger arrival time $t_{D_0}$
 at $x=0$.

 The observed ballistic dynamics of the de-correlator is in stark contrast to
 the purely diffusive relaxational dynamics of the spin correlators in the
 system.

 \sectionn{Outlook} %
 We have analysed the $T$-dependence of chaos in a classical many-body system
 with local interactions, centered around the butterfly effect, with
 implications for the physics of spin liquids, the classical-quantum
 correspondence for chaotic spin systems, the relevance of OTOC's in classical
 chaos, and the relation of microscopic chaos to macroscopic transport.
 
 % Specifically, we obtain the full temperature dependence of the (leading)
 % Lyapunov exponent $\mu$ and the butterfly speed $v_b$. At low temperaturs we
 % find power law behaviour, with $\mu \sim T^{0.48}$ and $v_b \sim T^{0.23}$.
 % The observed low temperature behaviour is consistent with the proposed
 % relation to the diffusion constant, $D \sim v_b^2/\mu$. Moreover, the
 % vanishing of the (leading) Lyapunov exponent is found to be parametrically
 % slower than that of the quantum bound $\mu \sim T$.
 
 % Somewhat similarly to the abovementioned quantum system, the SYK model, in
 % the present case, due to the complete absence of magnetic ordering, the high
 % $T$ properties survive in the correlated low temperature regime all the way
 % to $T=0$, while exhibiting interesting scaling with $T$. %However,
 % one major difference from an SYK-type model and its classical counterparts is
 % the notion of locality in real space of our lattice models, that allows a
 % notion of dimensionality and results in spatial structure of the butterfly
 % effect. Such a sense of spatial locality is ubiquitously present in condensed
 % matter systems.
   
 Many follow-on questions naturally pose themselves, e.g.\ concerning the role
 of phase transitions and order, the nature of the semiclassical limit, and the
 `transition' into an integrable regime with an increasing number of conserved
 quantities. Even at the classical level, small perturbations may lead to
 ordering and a reduction of the dimensionality of the ground state manifold. In
 such cases the observed effects are expected to survive above the concomitant
 ordering temperatures, which are often much smaller than the leading
 interaction scale, opening up a robust spin liquid regime .
   
 Another interesting aspect is the effect of the quantum fluctuations on this
 classical model. Such quantum fluctuations would, again, generally quench the
 ground state entropy. However, this may not necessarily lead to an ordered
 state, but a long range quantum entangled spin liquid state expected for this
 system in $S=1/2$ limit \cite{PhysRevB.65.224412}. The crossover to such a
 quantum coherent regime would then be accompanied by sharp signature in the
 indicators of many-body chaos studied above.

 % We studied in detail the temperature dependence of the butterfly effect in a
 % classical chaotic many-body system. Whereas the system itself shows purely
 % relaxational diffussive behaviour down to $T=0$ as captured by the spin-spin
 % auto-correlation function and the dynamical structure factor, an
 % appropriately defined out-of-time-order correlator reveals the chaotic nature
 % of the dynamics with a ballistically spreading wavefront and exponential
 % growth.

 \sectionn{Acknowledgements} This work was supported by the Max-Planck partner
 group on strongly correlated systems at ICTS, the Deutsche
 Forschungsgemeinschaft under SFB 1143 and SERB-DST (India) through project
 grant No. ECR/2017/000504. The authors acknowledge fruitful discussion and
 collaboration on related work with S. Banerjee, A. Dhar, A. Das, D. A. Huse, A.
 Kundu and S. S. Ray.

 \bibliography{butterfly.bib}{}

\begin{thebibliography}{70}%
\makeatletter
\providecommand \@ifxundefined [1]{%
 \@ifx{#1\undefined}
}%
\providecommand \@ifnum [1]{%
 \ifnum #1\expandafter \@firstoftwo
 \else \expandafter \@secondoftwo
 \fi
}%
\providecommand \@ifx [1]{%
 \ifx #1\expandafter \@firstoftwo
 \else \expandafter \@secondoftwo
 \fi
}%
\providecommand \natexlab [1]{#1}%
\providecommand \enquote  [1]{``#1''}%
\providecommand \bibnamefont  [1]{#1}%
\providecommand \bibfnamefont [1]{#1}%
\providecommand \citenamefont [1]{#1}%
\providecommand \href@noop [0]{\@secondoftwo}%
\providecommand \href [0]{\begingroup \@sanitize@url \@href}%
\providecommand \@href[1]{\@@startlink{#1}\@@href}%
\providecommand \@@href[1]{\endgroup#1\@@endlink}%
\providecommand \@sanitize@url [0]{\catcode `\\12\catcode `\$12\catcode
  `\&12\catcode `\#12\catcode `\^12\catcode `\_12\catcode `\%12\relax}%
\providecommand \@@startlink[1]{}%
\providecommand \@@endlink[0]{}%
\providecommand \url  [0]{\begingroup\@sanitize@url \@url }%
\providecommand \@url [1]{\endgroup\@href {#1}{\urlprefix }}%
\providecommand \urlprefix  [0]{URL }%
\providecommand \Eprint [0]{\href }%
\providecommand \doibase [0]{http://dx.doi.org/}%
\providecommand \selectlanguage [0]{\@gobble}%
\providecommand \bibinfo  [0]{\@secondoftwo}%
\providecommand \bibfield  [0]{\@secondoftwo}%
\providecommand \translation [1]{[#1]}%
\providecommand \BibitemOpen [0]{}%
\providecommand \bibitemStop [0]{}%
\providecommand \bibitemNoStop [0]{.\EOS\space}%
\providecommand \EOS [0]{\spacefactor3000\relax}%
\providecommand \BibitemShut  [1]{\csname bibitem#1\endcsname}%
\let\auto@bib@innerbib\@empty
%</preamble>
\bibitem [{\citenamefont {Dellago}\ and\ \citenamefont
  {Posch}(1997)}]{Dellago1997}%
  \BibitemOpen
  \bibfield  {author} {\bibinfo {author} {\bibfnamefont {C.}~\bibnamefont
  {Dellago}}\ and\ \bibinfo {author} {\bibfnamefont {H.}~\bibnamefont
  {Posch}},\ }\bibfield  {title} {\enquote {\bibinfo {title} {{Kolmogorov-Sinai
  entropy and Lyapunov spectra of a hard-sphere gas}},}\ }\href {\doibase
  10.1016/s0378-4371(97)00131-3} {\bibfield  {journal} {\bibinfo  {journal}
  {Physica A: Statistical Mechanics and its Applications}\ }\textbf {\bibinfo
  {volume} {240}},\ \bibinfo {pages} {68--83} (\bibinfo {year}
  {1997})}\BibitemShut {NoStop}%
\bibitem [{\citenamefont {Hoover}\ \emph {et~al.}(2002)\citenamefont {Hoover},
  \citenamefont {Posch}, \citenamefont {Forster}, \citenamefont {Dellago},\
  and\ \citenamefont {Zhou}}]{Hoover2002}%
  \BibitemOpen
  \bibfield  {author} {\bibinfo {author} {\bibfnamefont {W.~G.}\ \bibnamefont
  {Hoover}}, \bibinfo {author} {\bibfnamefont {H.~A.}\ \bibnamefont {Posch}},
  \bibinfo {author} {\bibfnamefont {C.}~\bibnamefont {Forster}}, \bibinfo
  {author} {\bibfnamefont {C.}~\bibnamefont {Dellago}}, \ and\ \bibinfo
  {author} {\bibfnamefont {M.}~\bibnamefont {Zhou}},\ }\bibfield  {title}
  {\enquote {\bibinfo {title} {{Lyapunov Modes of Two-Dimensional Many-Body
  Systems; Soft Disks, Hard Disks, and Rotors}},}\ }\href {\doibase
  10.1023/A:1020474901341} {\bibfield  {journal} {\bibinfo  {journal} {Journal
  of Statistical Physics}\ }\textbf {\bibinfo {volume} {109}},\ \bibinfo
  {pages} {765--776} (\bibinfo {year} {2002})}\BibitemShut {NoStop}%
\bibitem [{\citenamefont {Latora}\ \emph {et~al.}(1998)\citenamefont {Latora},
  \citenamefont {Rapisarda},\ and\ \citenamefont {Ruffo}}]{Latora1998}%
  \BibitemOpen
  \bibfield  {author} {\bibinfo {author} {\bibfnamefont {V.}~\bibnamefont
  {Latora}}, \bibinfo {author} {\bibfnamefont {A.}~\bibnamefont {Rapisarda}}, \
  and\ \bibinfo {author} {\bibfnamefont {S.}~\bibnamefont {Ruffo}},\ }\bibfield
   {title} {\enquote {\bibinfo {title} {{Lyapunov Instability and Finite Size
  Effects in a System with Long-Range Forces}},}\ }\href {\doibase
  10.1103/PhysRevLett.80.692} {\bibfield  {journal} {\bibinfo  {journal} {Phys.
  Rev. Lett.}\ }\textbf {\bibinfo {volume} {80}},\ \bibinfo {pages} {692--695}
  (\bibinfo {year} {1998})}\BibitemShut {NoStop}%
\bibitem [{\citenamefont {van Zon}\ \emph {et~al.}(1998)\citenamefont {van
  Zon}, \citenamefont {van Beijeren},\ and\ \citenamefont {Dellago}}]{Zon1998}%
  \BibitemOpen
  \bibfield  {author} {\bibinfo {author} {\bibfnamefont {R.}~\bibnamefont {van
  Zon}}, \bibinfo {author} {\bibfnamefont {H.}~\bibnamefont {van Beijeren}}, \
  and\ \bibinfo {author} {\bibfnamefont {C.}~\bibnamefont {Dellago}},\
  }\bibfield  {title} {\enquote {\bibinfo {title} {{Largest Lyapunov Exponent
  for Many Particle Systems at Low Densities}},}\ }\href {\doibase
  10.1103/PhysRevLett.80.2035} {\bibfield  {journal} {\bibinfo  {journal}
  {Phys. Rev. Lett.}\ }\textbf {\bibinfo {volume} {80}},\ \bibinfo {pages}
  {2035--2038} (\bibinfo {year} {1998})}\BibitemShut {NoStop}%
\bibitem [{\citenamefont {van Beijeren}\ \emph {et~al.}(1997)\citenamefont {van
  Beijeren}, \citenamefont {Dorfman}, \citenamefont {Posch},\ and\
  \citenamefont {Dellago}}]{Beijeren1997}%
  \BibitemOpen
  \bibfield  {author} {\bibinfo {author} {\bibfnamefont {H.}~\bibnamefont {van
  Beijeren}}, \bibinfo {author} {\bibfnamefont {J.~R.}\ \bibnamefont
  {Dorfman}}, \bibinfo {author} {\bibfnamefont {H.~A.}\ \bibnamefont {Posch}},
  \ and\ \bibinfo {author} {\bibfnamefont {C.}~\bibnamefont {Dellago}},\
  }\bibfield  {title} {\enquote {\bibinfo {title} {{Kolmogorov-Sinai entropy
  for dilute gases in equilibrium}},}\ }\href {\doibase
  10.1103/PhysRevE.56.5272} {\bibfield  {journal} {\bibinfo  {journal} {Phys.
  Rev. E}\ }\textbf {\bibinfo {volume} {56}},\ \bibinfo {pages} {5272--5277}
  (\bibinfo {year} {1997})}\BibitemShut {NoStop}%
\bibitem [{\citenamefont {de~Wijn}\ and\ \citenamefont {van
  Beijeren}(2004)}]{Wijn2004}%
  \BibitemOpen
  \bibfield  {author} {\bibinfo {author} {\bibfnamefont {A.~S.}\ \bibnamefont
  {de~Wijn}}\ and\ \bibinfo {author} {\bibfnamefont {H.}~\bibnamefont {van
  Beijeren}},\ }\bibfield  {title} {\enquote {\bibinfo {title} {{Goldstone
  modes in Lyapunov spectra of hard sphere systems}},}\ }\href {\doibase
  10.1103/PhysRevE.70.016207} {\bibfield  {journal} {\bibinfo  {journal} {Phys.
  Rev. E}\ }\textbf {\bibinfo {volume} {70}},\ \bibinfo {pages} {016207}
  (\bibinfo {year} {2004})}\BibitemShut {NoStop}%
\bibitem [{\citenamefont {de~Wijn}(2005)}]{Wijn2005}%
  \BibitemOpen
  \bibfield  {author} {\bibinfo {author} {\bibfnamefont {A.~S.}\ \bibnamefont
  {de~Wijn}},\ }\bibfield  {title} {\enquote {\bibinfo {title} {{Lyapunov
  spectra of billiards with cylindrical scatterers: Comparison with
  many-particle systems}},}\ }\href {\doibase 10.1103/PhysRevE.72.026216}
  {\bibfield  {journal} {\bibinfo  {journal} {Phys. Rev. E}\ }\textbf {\bibinfo
  {volume} {72}},\ \bibinfo {pages} {026216} (\bibinfo {year}
  {2005})}\BibitemShut {NoStop}%
\bibitem [{\citenamefont {de~Wijn}\ \emph {et~al.}(2012)\citenamefont
  {de~Wijn}, \citenamefont {Hess},\ and\ \citenamefont {Fine}}]{Wijn2012}%
  \BibitemOpen
  \bibfield  {author} {\bibinfo {author} {\bibfnamefont {A.~S.}\ \bibnamefont
  {de~Wijn}}, \bibinfo {author} {\bibfnamefont {B.}~\bibnamefont {Hess}}, \
  and\ \bibinfo {author} {\bibfnamefont {B.~V.}\ \bibnamefont {Fine}},\
  }\bibfield  {title} {\enquote {\bibinfo {title} {{Largest Lyapunov Exponents
  for Lattices of Interacting Classical Spins}},}\ }\href {\doibase
  10.1103/PhysRevLett.109.034101} {\bibfield  {journal} {\bibinfo  {journal}
  {Phys. Rev. Lett.}\ }\textbf {\bibinfo {volume} {109}},\ \bibinfo {pages}
  {034101} (\bibinfo {year} {2012})}\BibitemShut {NoStop}%
\bibitem [{\citenamefont {de~Wijn}\ \emph {et~al.}(2013)\citenamefont
  {de~Wijn}, \citenamefont {Hess},\ and\ \citenamefont {Fine}}]{Wijn2013}%
  \BibitemOpen
  \bibfield  {author} {\bibinfo {author} {\bibfnamefont {A.~S.}\ \bibnamefont
  {de~Wijn}}, \bibinfo {author} {\bibfnamefont {B.}~\bibnamefont {Hess}}, \
  and\ \bibinfo {author} {\bibfnamefont {B.~V.}\ \bibnamefont {Fine}},\
  }\bibfield  {title} {\enquote {\bibinfo {title} {{Lyapunov instabilities in
  lattices of interacting classical spins at infinite temperature}},}\ }\href
  {\doibase 10.1088/1751-8113/46/25/254012} {\bibfield  {journal} {\bibinfo
  {journal} {Journal of Physics A: Mathematical and Theoretical}\ }\textbf
  {\bibinfo {volume} {46}},\ \bibinfo {pages} {254012} (\bibinfo {year}
  {2013})}\BibitemShut {NoStop}%
\bibitem [{\citenamefont {Gaspard}\ \emph {et~al.}(1998)\citenamefont
  {Gaspard}, \citenamefont {Briggs}, \citenamefont {Francis}, \citenamefont
  {Sengers}, \citenamefont {Gammon}, \citenamefont {Dorfman},\ and\
  \citenamefont {Calabrese}}]{gaspard1998experimental}%
  \BibitemOpen
  \bibfield  {author} {\bibinfo {author} {\bibfnamefont {P.}~\bibnamefont
  {Gaspard}}, \bibinfo {author} {\bibfnamefont {M.~E.}\ \bibnamefont {Briggs}},
  \bibinfo {author} {\bibfnamefont {M.~K.}\ \bibnamefont {Francis}}, \bibinfo
  {author} {\bibfnamefont {J.~V.}\ \bibnamefont {Sengers}}, \bibinfo {author}
  {\bibfnamefont {R.~W.}\ \bibnamefont {Gammon}}, \bibinfo {author}
  {\bibfnamefont {J.~R.}\ \bibnamefont {Dorfman}}, \ and\ \bibinfo {author}
  {\bibfnamefont {R.~V.}\ \bibnamefont {Calabrese}},\ }\bibfield  {title}
  {\enquote {\bibinfo {title} {{Experimental evidence for microscopic
  chaos}},}\ }\href {\doibase 10.1038/29721} {\bibfield  {journal} {\bibinfo
  {journal} {Nature}\ }\textbf {\bibinfo {volume} {394}},\ \bibinfo {pages}
  {865--868} (\bibinfo {year} {1998})}\BibitemShut {NoStop}%
\bibitem [{\citenamefont {Gaspard}\ \emph {et~al.}(1999)\citenamefont
  {Gaspard}, \citenamefont {Briggs}, \citenamefont {Francis}, \citenamefont
  {Sengers}, \citenamefont {Gammon}, \citenamefont {Dorfman},\ and\
  \citenamefont {Calabrese}}]{dettmann1999statistical}%
  \BibitemOpen
  \bibfield  {author} {\bibinfo {author} {\bibfnamefont {P.}~\bibnamefont
  {Gaspard}}, \bibinfo {author} {\bibfnamefont {M.~E.}\ \bibnamefont {Briggs}},
  \bibinfo {author} {\bibfnamefont {M.~K.}\ \bibnamefont {Francis}}, \bibinfo
  {author} {\bibfnamefont {J.~V.}\ \bibnamefont {Sengers}}, \bibinfo {author}
  {\bibfnamefont {R.~W.}\ \bibnamefont {Gammon}}, \bibinfo {author}
  {\bibfnamefont {J.~R.}\ \bibnamefont {Dorfman}}, \ and\ \bibinfo {author}
  {\bibfnamefont {R.~V.}\ \bibnamefont {Calabrese}},\ }\bibfield  {title}
  {\enquote {\bibinfo {title} {{Microscopic chaos from brownian motion?}}}\
  }\href {\doibase 10.1038/44764} {\bibfield  {journal} {\bibinfo  {journal}
  {Nature}\ }\textbf {\bibinfo {volume} {401}},\ \bibinfo {pages} {876--876}
  (\bibinfo {year} {1999})}\BibitemShut {NoStop}%
\bibitem [{\citenamefont {Lorenz}(1963)}]{lorenz1963deterministic}%
  \BibitemOpen
  \bibfield  {author} {\bibinfo {author} {\bibfnamefont {E.~N.}\ \bibnamefont
  {Lorenz}},\ }\bibfield  {title} {\enquote {\bibinfo {title} {{Deterministic
  nonperiodic flow}},}\ }\href
  {http://journals.ametsoc.org/doi/abs/10.1175/1520-0469(1963)020%3C0130:DNF%3E2.0.CO;2}
  {\bibfield  {journal} {\bibinfo  {journal} {Journal of the atmospheric
  sciences}\ }\textbf {\bibinfo {volume} {20}},\ \bibinfo {pages} {130--141}
  (\bibinfo {year} {1963})}\BibitemShut {NoStop}%
\bibitem [{\citenamefont {Lorenz}(1993)}]{lorenz1996essence}%
  \BibitemOpen
  \bibfield  {author} {\bibinfo {author} {\bibfnamefont {E.~N.}\ \bibnamefont
  {Lorenz}},\ }\href
  {http://www.washington.edu/uwpress/search/books/LORESS.html} {\emph {\bibinfo
  {title} {{The essence of chaos}}}}\ (\bibinfo  {publisher} {University of
  Washington Press, Seattle, Washington},\ \bibinfo {year} {1993})\BibitemShut
  {NoStop}%
\bibitem [{\citenamefont {Lorenz}(2001)}]{lorenz2000butterfly}%
  \BibitemOpen
  \bibfield  {author} {\bibinfo {author} {\bibfnamefont {E.}~\bibnamefont
  {Lorenz}},\ }\bibfield  {title} {\enquote {\bibinfo {title} {The butterfly
  effect},}\ }in\ \href {\doibase 10.1142/9789812386472_0007} {\emph {\bibinfo
  {booktitle} {The Chaos Avant-Garde}}}\ (\bibinfo  {publisher} {{World}
  {Scientific}},\ \bibinfo {year} {2001})\ pp.\ \bibinfo {pages}
  {91--94}\BibitemShut {NoStop}%
\bibitem [{\citenamefont {Hilborn}(2004)}]{hilborn2004sea}%
  \BibitemOpen
  \bibfield  {author} {\bibinfo {author} {\bibfnamefont {R.~C.}\ \bibnamefont
  {Hilborn}},\ }\bibfield  {title} {\enquote {\bibinfo {title} {{Sea gulls,
  butterflies, and grasshoppers: A brief history of the butterfly effect in
  nonlinear dynamics}},}\ }\href
  {http://aapt.scitation.org/doi/abs/10.1119/1.1636492} {\bibfield  {journal}
  {\bibinfo  {journal} {American Journal of Physics}\ }\textbf {\bibinfo
  {volume} {72}},\ \bibinfo {pages} {425--427} (\bibinfo {year}
  {2004})}\BibitemShut {NoStop}%
\bibitem [{\citenamefont {Maldacena}\ \emph {et~al.}(2016)\citenamefont
  {Maldacena}, \citenamefont {Shenker},\ and\ \citenamefont
  {Stanford}}]{Maldacena2016}%
  \BibitemOpen
  \bibfield  {author} {\bibinfo {author} {\bibfnamefont {J.}~\bibnamefont
  {Maldacena}}, \bibinfo {author} {\bibfnamefont {S.~H.}\ \bibnamefont
  {Shenker}}, \ and\ \bibinfo {author} {\bibfnamefont {D.}~\bibnamefont
  {Stanford}},\ }\bibfield  {title} {\enquote {\bibinfo {title} {{A bound on
  chaos}},}\ }\href {\doibase 10.1007/JHEP08(2016)106} {\bibfield  {journal}
  {\bibinfo  {journal} {Journal of High Energy Physics}\ }\textbf {\bibinfo
  {volume} {2016}},\ \bibinfo {pages} {106} (\bibinfo {year}
  {2016})}\BibitemShut {NoStop}%
\bibitem [{\citenamefont {Hosur}\ \emph {et~al.}(2016)\citenamefont {Hosur},
  \citenamefont {Qi}, \citenamefont {Roberts},\ and\ \citenamefont
  {Yoshida}}]{Hosur2016}%
  \BibitemOpen
  \bibfield  {author} {\bibinfo {author} {\bibfnamefont {P.}~\bibnamefont
  {Hosur}}, \bibinfo {author} {\bibfnamefont {X.-L.}\ \bibnamefont {Qi}},
  \bibinfo {author} {\bibfnamefont {D.~A.}\ \bibnamefont {Roberts}}, \ and\
  \bibinfo {author} {\bibfnamefont {B.}~\bibnamefont {Yoshida}},\ }\bibfield
  {title} {\enquote {\bibinfo {title} {{Chaos in quantum channels}},}\ }\href
  {\doibase 10.1007/JHEP02(2016)004} {\bibfield  {journal} {\bibinfo  {journal}
  {Journal of High Energy Physics}\ }\textbf {\bibinfo {volume} {2016}},\
  \bibinfo {pages} {4} (\bibinfo {year} {2016})}\BibitemShut {NoStop}%
\bibitem [{\citenamefont {Roberts}\ \emph {et~al.}(2015)\citenamefont
  {Roberts}, \citenamefont {Stanford},\ and\ \citenamefont
  {Susskind}}]{Roberts2015}%
  \BibitemOpen
  \bibfield  {author} {\bibinfo {author} {\bibfnamefont {D.~A.}\ \bibnamefont
  {Roberts}}, \bibinfo {author} {\bibfnamefont {D.}~\bibnamefont {Stanford}}, \
  and\ \bibinfo {author} {\bibfnamefont {L.}~\bibnamefont {Susskind}},\
  }\bibfield  {title} {\enquote {\bibinfo {title} {{Localized shocks}},}\
  }\href {\doibase 10.1007/JHEP03(2015)051} {\bibfield  {journal} {\bibinfo
  {journal} {Journal of High Energy Physics}\ }\textbf {\bibinfo {volume}
  {2015}},\ \bibinfo {pages} {51} (\bibinfo {year} {2015})}\BibitemShut
  {NoStop}%
\bibitem [{\citenamefont {Roberts}\ and\ \citenamefont
  {Stanford}(2015)}]{Roberts2015a}%
  \BibitemOpen
  \bibfield  {author} {\bibinfo {author} {\bibfnamefont {D.~A.}\ \bibnamefont
  {Roberts}}\ and\ \bibinfo {author} {\bibfnamefont {D.}~\bibnamefont
  {Stanford}},\ }\bibfield  {title} {\enquote {\bibinfo {title} {{Diagnosing
  Chaos Using Four-Point Functions in Two-Dimensional Conformal Field
  Theory}},}\ }\href {\doibase 10.1103/PhysRevLett.115.131603} {\bibfield
  {journal} {\bibinfo  {journal} {Phys. Rev. Lett.}\ }\textbf {\bibinfo
  {volume} {115}},\ \bibinfo {pages} {131603} (\bibinfo {year}
  {2015})}\BibitemShut {NoStop}%
\bibitem [{\citenamefont {Shenker}\ and\ \citenamefont
  {Stanford}(2014)}]{Shenker2014}%
  \BibitemOpen
  \bibfield  {author} {\bibinfo {author} {\bibfnamefont {S.~H.}\ \bibnamefont
  {Shenker}}\ and\ \bibinfo {author} {\bibfnamefont {D.}~\bibnamefont
  {Stanford}},\ }\bibfield  {title} {\enquote {\bibinfo {title} {{Black holes
  and the butterfly effect}},}\ }\href {\doibase 10.1007/JHEP03(2014)067}
  {\bibfield  {journal} {\bibinfo  {journal} {Journal of High Energy Physics}\
  }\textbf {\bibinfo {volume} {2014}},\ \bibinfo {pages} {67} (\bibinfo {year}
  {2014})}\BibitemShut {NoStop}%
\bibitem [{\citenamefont {Cotler}\ \emph {et~al.}(2017)\citenamefont {Cotler},
  \citenamefont {Gur-Ari}, \citenamefont {Hanada}, \citenamefont {Polchinski},
  \citenamefont {Saad}, \citenamefont {Shenker}, \citenamefont {Stanford},
  \citenamefont {Streicher},\ and\ \citenamefont {Tezuka}}]{Cotler2017}%
  \BibitemOpen
  \bibfield  {author} {\bibinfo {author} {\bibfnamefont {J.~S.}\ \bibnamefont
  {Cotler}}, \bibinfo {author} {\bibfnamefont {G.}~\bibnamefont {Gur-Ari}},
  \bibinfo {author} {\bibfnamefont {M.}~\bibnamefont {Hanada}}, \bibinfo
  {author} {\bibfnamefont {J.}~\bibnamefont {Polchinski}}, \bibinfo {author}
  {\bibfnamefont {P.}~\bibnamefont {Saad}}, \bibinfo {author} {\bibfnamefont
  {S.~H.}\ \bibnamefont {Shenker}}, \bibinfo {author} {\bibfnamefont
  {D.}~\bibnamefont {Stanford}}, \bibinfo {author} {\bibfnamefont
  {A.}~\bibnamefont {Streicher}}, \ and\ \bibinfo {author} {\bibfnamefont
  {M.}~\bibnamefont {Tezuka}},\ }\bibfield  {title} {\enquote {\bibinfo {title}
  {{Black holes and random matrices}},}\ }\href {\doibase
  10.1007/JHEP05(2017)118} {\bibfield  {journal} {\bibinfo  {journal} {Journal
  of High Energy Physics}\ }\textbf {\bibinfo {volume} {2017}},\ \bibinfo
  {pages} {118} (\bibinfo {year} {2017})}\BibitemShut {NoStop}%
\bibitem [{\citenamefont {D\'ora}\ and\ \citenamefont
  {Moessner}(2017)}]{Dora2017}%
  \BibitemOpen
  \bibfield  {author} {\bibinfo {author} {\bibfnamefont {B.}~\bibnamefont
  {D\'ora}}\ and\ \bibinfo {author} {\bibfnamefont {R.}~\bibnamefont
  {Moessner}},\ }\bibfield  {title} {\enquote {\bibinfo {title}
  {{Out-of-Time-Ordered Density Correlators in Luttinger Liquids}},}\ }\href
  {\doibase 10.1103/PhysRevLett.119.026802} {\bibfield  {journal} {\bibinfo
  {journal} {Phys. Rev. Lett.}\ }\textbf {\bibinfo {volume} {119}},\ \bibinfo
  {pages} {026802} (\bibinfo {year} {2017})}\BibitemShut {NoStop}%
\bibitem [{\citenamefont {Menezes}\ and\ \citenamefont
  {Marino}(2018)}]{Menezes2018}%
  \BibitemOpen
  \bibfield  {author} {\bibinfo {author} {\bibfnamefont {G.}~\bibnamefont
  {Menezes}}\ and\ \bibinfo {author} {\bibfnamefont {J.}~\bibnamefont
  {Marino}},\ }\bibfield  {title} {\enquote {\bibinfo {title} {Slow scrambling
  in sonic black holes},}\ }\href {\doibase 10.1209/0295-5075/121/60002}
  {\bibfield  {journal} {\bibinfo  {journal} {{EPL} (Europhysics Letters)}\
  }\textbf {\bibinfo {volume} {121}},\ \bibinfo {pages} {60002} (\bibinfo
  {year} {2018})}\BibitemShut {NoStop}%
\bibitem [{\citenamefont {Rozenbaum}\ \emph {et~al.}(2017)\citenamefont
  {Rozenbaum}, \citenamefont {Ganeshan},\ and\ \citenamefont
  {Galitski}}]{Rozenbaum2017}%
  \BibitemOpen
  \bibfield  {author} {\bibinfo {author} {\bibfnamefont {E.~B.}\ \bibnamefont
  {Rozenbaum}}, \bibinfo {author} {\bibfnamefont {S.}~\bibnamefont {Ganeshan}},
  \ and\ \bibinfo {author} {\bibfnamefont {V.}~\bibnamefont {Galitski}},\
  }\bibfield  {title} {\enquote {\bibinfo {title} {{Lyapunov Exponent and
  Out-of-Time-Ordered Correlator's Growth Rate in a Chaotic System}},}\ }\href
  {\doibase 10.1103/PhysRevLett.118.086801} {\bibfield  {journal} {\bibinfo
  {journal} {Phys. Rev. Lett.}\ }\textbf {\bibinfo {volume} {118}},\ \bibinfo
  {pages} {086801} (\bibinfo {year} {2017})}\BibitemShut {NoStop}%
\bibitem [{\citenamefont {Luitz}\ and\ \citenamefont
  {Bar~Lev}(2017)}]{Luitz2017}%
  \BibitemOpen
  \bibfield  {author} {\bibinfo {author} {\bibfnamefont {D.~J.}\ \bibnamefont
  {Luitz}}\ and\ \bibinfo {author} {\bibfnamefont {Y.}~\bibnamefont
  {Bar~Lev}},\ }\bibfield  {title} {\enquote {\bibinfo {title} {{Information
  propagation in isolated quantum systems}},}\ }\href {\doibase
  10.1103/PhysRevB.96.020406} {\bibfield  {journal} {\bibinfo  {journal} {Phys.
  Rev. B}\ }\textbf {\bibinfo {volume} {96}},\ \bibinfo {pages} {020406}
  (\bibinfo {year} {2017})}\BibitemShut {NoStop}%
\bibitem [{\citenamefont {Kukuljan}\ \emph {et~al.}(2017)\citenamefont
  {Kukuljan}, \citenamefont {Grozdanov},\ and\ \citenamefont
  {Prosen}}]{Kukuljan2017}%
  \BibitemOpen
  \bibfield  {author} {\bibinfo {author} {\bibfnamefont {I.}~\bibnamefont
  {Kukuljan}}, \bibinfo {author} {\bibfnamefont {S.~c.~v.}\ \bibnamefont
  {Grozdanov}}, \ and\ \bibinfo {author} {\bibfnamefont {T.~c.~v.}\
  \bibnamefont {Prosen}},\ }\bibfield  {title} {\enquote {\bibinfo {title}
  {{Weak quantum chaos}},}\ }\href {\doibase 10.1103/PhysRevB.96.060301}
  {\bibfield  {journal} {\bibinfo  {journal} {Phys. Rev. B}\ }\textbf {\bibinfo
  {volume} {96}},\ \bibinfo {pages} {060301} (\bibinfo {year}
  {2017})}\BibitemShut {NoStop}%
\bibitem [{\citenamefont {Aleiner}\ \emph {et~al.}(2016)\citenamefont
  {Aleiner}, \citenamefont {Faoro},\ and\ \citenamefont {Ioffe}}]{Aleiner2016}%
  \BibitemOpen
  \bibfield  {author} {\bibinfo {author} {\bibfnamefont {I.~L.}\ \bibnamefont
  {Aleiner}}, \bibinfo {author} {\bibfnamefont {L.}~\bibnamefont {Faoro}}, \
  and\ \bibinfo {author} {\bibfnamefont {L.~B.}\ \bibnamefont {Ioffe}},\
  }\bibfield  {title} {\enquote {\bibinfo {title} {{Microscopic model of
  quantum butterfly effect: Out-of-time-order correlators and traveling
  combustion waves}},}\ }\href {\doibase 10.1016/j.aop.2016.09.006} {\bibfield
  {journal} {\bibinfo  {journal} {Annals of Physics}\ }\textbf {\bibinfo
  {volume} {375}},\ \bibinfo {pages} {378--406} (\bibinfo {year}
  {2016})}\BibitemShut {NoStop}%
\bibitem [{\citenamefont {Lin}\ and\ \citenamefont
  {Motrunich}(2018)}]{Lin2018}%
  \BibitemOpen
  \bibfield  {author} {\bibinfo {author} {\bibfnamefont {C.-J.}\ \bibnamefont
  {Lin}}\ and\ \bibinfo {author} {\bibfnamefont {O.~I.}\ \bibnamefont
  {Motrunich}},\ }\bibfield  {title} {\enquote {\bibinfo {title}
  {{Out-of-time-ordered correlators in a quantum Ising chain}},}\ }\href
  {\doibase 10.1103/PhysRevB.97.144304} {\bibfield  {journal} {\bibinfo
  {journal} {Phys. Rev. B}\ }\textbf {\bibinfo {volume} {97}},\ \bibinfo
  {pages} {144304} (\bibinfo {year} {2018})}\BibitemShut {NoStop}%
\bibitem [{\citenamefont {Chen}\ \emph {et~al.}(2016)\citenamefont {Chen},
  \citenamefont {Zhou}, \citenamefont {Huse},\ and\ \citenamefont
  {Fradkin}}]{Chen2016}%
  \BibitemOpen
  \bibfield  {author} {\bibinfo {author} {\bibfnamefont {X.}~\bibnamefont
  {Chen}}, \bibinfo {author} {\bibfnamefont {T.}~\bibnamefont {Zhou}}, \bibinfo
  {author} {\bibfnamefont {D.~A.}\ \bibnamefont {Huse}}, \ and\ \bibinfo
  {author} {\bibfnamefont {E.}~\bibnamefont {Fradkin}},\ }\bibfield  {title}
  {\enquote {\bibinfo {title} {{Out-of-time-order correlations in many-body
  localized and thermal phases}},}\ }\href {\doibase 10.1002/andp.201600332}
  {\bibfield  {journal} {\bibinfo  {journal} {Annalen der Physik}\ }\textbf
  {\bibinfo {volume} {529}},\ \bibinfo {pages} {1600332} (\bibinfo {year}
  {2016})}\BibitemShut {NoStop}%
\bibitem [{\citenamefont {{Chan}}\ \emph {et~al.}(2017)\citenamefont {{Chan}},
  \citenamefont {{De Luca}},\ and\ \citenamefont {{Chalker}}}]{Chan2017}%
  \BibitemOpen
  \bibfield  {author} {\bibinfo {author} {\bibfnamefont {A.}~\bibnamefont
  {{Chan}}}, \bibinfo {author} {\bibfnamefont {A.}~\bibnamefont {{De Luca}}}, \
  and\ \bibinfo {author} {\bibfnamefont {J.~T.}\ \bibnamefont {{Chalker}}},\
  }\bibfield  {title} {\enquote {\bibinfo {title} {{Solution of a minimal model
  for many-body quantum chaos}},}\ }\href@noop {} {\bibfield  {journal}
  {\bibinfo  {journal} {ArXiv e-prints}\ } (\bibinfo {year} {2017})},\ \Eprint
  {http://arxiv.org/abs/1712.06836} {arXiv:1712.06836 [cond-mat.stat-mech]}
  \BibitemShut {NoStop}%
\bibitem [{\citenamefont {Nahum}\ \emph {et~al.}(2018)\citenamefont {Nahum},
  \citenamefont {Vijay},\ and\ \citenamefont {Haah}}]{Nahum2018}%
  \BibitemOpen
  \bibfield  {author} {\bibinfo {author} {\bibfnamefont {A.}~\bibnamefont
  {Nahum}}, \bibinfo {author} {\bibfnamefont {S.}~\bibnamefont {Vijay}}, \ and\
  \bibinfo {author} {\bibfnamefont {J.}~\bibnamefont {Haah}},\ }\bibfield
  {title} {\enquote {\bibinfo {title} {{Operator Spreading in Random Unitary
  Circuits}},}\ }\href {\doibase 10.1103/PhysRevX.8.021014} {\bibfield
  {journal} {\bibinfo  {journal} {Phys. Rev. X}\ }\textbf {\bibinfo {volume}
  {8}},\ \bibinfo {pages} {021014} (\bibinfo {year} {2018})}\BibitemShut
  {NoStop}%
\bibitem [{\citenamefont {von Keyserlingk}\ \emph {et~al.}(2018)\citenamefont
  {von Keyserlingk}, \citenamefont {Rakovszky}, \citenamefont {Pollmann},\ and\
  \citenamefont {Sondhi}}]{Keyserlingk2018}%
  \BibitemOpen
  \bibfield  {author} {\bibinfo {author} {\bibfnamefont {C.~W.}\ \bibnamefont
  {von Keyserlingk}}, \bibinfo {author} {\bibfnamefont {T.}~\bibnamefont
  {Rakovszky}}, \bibinfo {author} {\bibfnamefont {F.}~\bibnamefont {Pollmann}},
  \ and\ \bibinfo {author} {\bibfnamefont {S.~L.}\ \bibnamefont {Sondhi}},\
  }\bibfield  {title} {\enquote {\bibinfo {title} {{Operator Hydrodynamics,
  OTOCs, and Entanglement Growth in Systems without Conservation Laws}},}\
  }\href {\doibase 10.1103/PhysRevX.8.021013} {\bibfield  {journal} {\bibinfo
  {journal} {Phys. Rev. X}\ }\textbf {\bibinfo {volume} {8}},\ \bibinfo {pages}
  {021013} (\bibinfo {year} {2018})}\BibitemShut {NoStop}%
\bibitem [{\citenamefont {{Rakovszky}}\ \emph {et~al.}(2017)\citenamefont
  {{Rakovszky}}, \citenamefont {{Pollmann}},\ and\ \citenamefont {{von
  Keyserlingk}}}]{Rakovszky2017}%
  \BibitemOpen
  \bibfield  {author} {\bibinfo {author} {\bibfnamefont {T.}~\bibnamefont
  {{Rakovszky}}}, \bibinfo {author} {\bibfnamefont {F.}~\bibnamefont
  {{Pollmann}}}, \ and\ \bibinfo {author} {\bibfnamefont {C.~W.}\ \bibnamefont
  {{von Keyserlingk}}},\ }\bibfield  {title} {\enquote {\bibinfo {title}
  {{Diffusive hydrodynamics of out-of-time-ordered correlators with charge
  conservation}},}\ }\href@noop {} {\bibfield  {journal} {\bibinfo  {journal}
  {ArXiv e-prints}\ } (\bibinfo {year} {2017})},\ \Eprint
  {http://arxiv.org/abs/1710.09827} {arXiv:1710.09827 [cond-mat.stat-mech]}
  \BibitemShut {NoStop}%
\bibitem [{\citenamefont {{Khemani}}\ \emph {et~al.}(2017)\citenamefont
  {{Khemani}}, \citenamefont {{Vishwanath}},\ and\ \citenamefont
  {{Huse}}}]{Khemani2017}%
  \BibitemOpen
  \bibfield  {author} {\bibinfo {author} {\bibfnamefont {V.}~\bibnamefont
  {{Khemani}}}, \bibinfo {author} {\bibfnamefont {A.}~\bibnamefont
  {{Vishwanath}}}, \ and\ \bibinfo {author} {\bibfnamefont {D.~A.}\
  \bibnamefont {{Huse}}},\ }\bibfield  {title} {\enquote {\bibinfo {title}
  {{Operator spreading and the emergence of dissipation in unitary dynamics
  with conservation laws}},}\ }\href@noop {} {\bibfield  {journal} {\bibinfo
  {journal} {ArXiv e-prints}\ } (\bibinfo {year} {2017})},\ \Eprint
  {http://arxiv.org/abs/1710.09835} {arXiv:1710.09835 [cond-mat.stat-mech]}
  \BibitemShut {NoStop}%
\bibitem [{\citenamefont {Stanford}(2016)}]{Stanford2016}%
  \BibitemOpen
  \bibfield  {author} {\bibinfo {author} {\bibfnamefont {D.}~\bibnamefont
  {Stanford}},\ }\bibfield  {title} {\enquote {\bibinfo {title} {{Many-body
  chaos at weak coupling}},}\ }\href {\doibase 10.1007/JHEP10(2016)009}
  {\bibfield  {journal} {\bibinfo  {journal} {Journal of High Energy Physics}\
  }\textbf {\bibinfo {volume} {2016}},\ \bibinfo {pages} {9} (\bibinfo {year}
  {2016})}\BibitemShut {NoStop}%
\bibitem [{\citenamefont {Patel}\ \emph {et~al.}(2017)\citenamefont {Patel},
  \citenamefont {Chowdhury}, \citenamefont {Sachdev},\ and\ \citenamefont
  {Swingle}}]{Patel2017}%
  \BibitemOpen
  \bibfield  {author} {\bibinfo {author} {\bibfnamefont {A.~A.}\ \bibnamefont
  {Patel}}, \bibinfo {author} {\bibfnamefont {D.}~\bibnamefont {Chowdhury}},
  \bibinfo {author} {\bibfnamefont {S.}~\bibnamefont {Sachdev}}, \ and\
  \bibinfo {author} {\bibfnamefont {B.}~\bibnamefont {Swingle}},\ }\bibfield
  {title} {\enquote {\bibinfo {title} {{Quantum Butterfly Effect in Weakly
  Interacting Diffusive Metals}},}\ }\href {\doibase 10.1103/PhysRevX.7.031047}
  {\bibfield  {journal} {\bibinfo  {journal} {Phys. Rev. X}\ }\textbf {\bibinfo
  {volume} {7}},\ \bibinfo {pages} {031047} (\bibinfo {year}
  {2017})}\BibitemShut {NoStop}%
\bibitem [{\citenamefont {Chowdhury}\ and\ \citenamefont
  {Swingle}(2017)}]{Chowdhury2017}%
  \BibitemOpen
  \bibfield  {author} {\bibinfo {author} {\bibfnamefont {D.}~\bibnamefont
  {Chowdhury}}\ and\ \bibinfo {author} {\bibfnamefont {B.}~\bibnamefont
  {Swingle}},\ }\bibfield  {title} {\enquote {\bibinfo {title} {{Onset of
  many-body chaos in the $O(N)$ model}},}\ }\href {\doibase
  10.1103/PhysRevD.96.065005} {\bibfield  {journal} {\bibinfo  {journal} {Phys.
  Rev. D}\ }\textbf {\bibinfo {volume} {96}},\ \bibinfo {pages} {065005}
  (\bibinfo {year} {2017})}\BibitemShut {NoStop}%
\bibitem [{\citenamefont {Srednicki}(1994)}]{Srednicki1994}%
  \BibitemOpen
  \bibfield  {author} {\bibinfo {author} {\bibfnamefont {M.}~\bibnamefont
  {Srednicki}},\ }\bibfield  {title} {\enquote {\bibinfo {title} {{Chaos and
  quantum thermalization}},}\ }\href {\doibase 10.1103/PhysRevE.50.888}
  {\bibfield  {journal} {\bibinfo  {journal} {Phys. Rev. E}\ }\textbf {\bibinfo
  {volume} {50}},\ \bibinfo {pages} {888--901} (\bibinfo {year}
  {1994})}\BibitemShut {NoStop}%
\bibitem [{\citenamefont {Gutzwiller}(1991)}]{Gutzwiller1991}%
  \BibitemOpen
  \bibfield  {author} {\bibinfo {author} {\bibfnamefont {M.~C.}\ \bibnamefont
  {Gutzwiller}},\ }\href {https://books.google.de/books?id=fnO3XYYpU54C} {\emph
  {\bibinfo {title} {{Chaos in Classical and Quantum Mechanics}}}},\
  Interdisciplinary Applied Mathematics\ (\bibinfo  {publisher} {Springer New
  York},\ \bibinfo {year} {1991})\BibitemShut {NoStop}%
\bibitem [{\citenamefont {{Larkin}}\ and\ \citenamefont
  {{Ovchinnikov}}(1969)}]{Larkin1969}%
  \BibitemOpen
  \bibfield  {author} {\bibinfo {author} {\bibfnamefont {A.~I.}\ \bibnamefont
  {{Larkin}}}\ and\ \bibinfo {author} {\bibfnamefont {Y.~N.}\ \bibnamefont
  {{Ovchinnikov}}},\ }\bibfield  {title} {\enquote {\bibinfo {title}
  {{Quasiclassical Method in the Theory of Superconductivity}},}\ }\href@noop
  {} {\bibfield  {journal} {\bibinfo  {journal} {Soviet Journal of Experimental
  and Theoretical Physics}\ }\textbf {\bibinfo {volume} {28}},\ \bibinfo
  {pages} {1200} (\bibinfo {year} {1969})}\BibitemShut {NoStop}%
\bibitem [{\citenamefont {Das}\ \emph {et~al.}(2018)\citenamefont {Das},
  \citenamefont {Chakrabarty}, \citenamefont {Dhar}, \citenamefont {Kundu},
  \citenamefont {Huse}, \citenamefont {Moessner}, \citenamefont {Ray},\ and\
  \citenamefont {Bhattacharjee}}]{Das2018}%
  \BibitemOpen
  \bibfield  {author} {\bibinfo {author} {\bibfnamefont {A.}~\bibnamefont
  {Das}}, \bibinfo {author} {\bibfnamefont {S.}~\bibnamefont {Chakrabarty}},
  \bibinfo {author} {\bibfnamefont {A.}~\bibnamefont {Dhar}}, \bibinfo {author}
  {\bibfnamefont {A.}~\bibnamefont {Kundu}}, \bibinfo {author} {\bibfnamefont
  {D.~A.}\ \bibnamefont {Huse}}, \bibinfo {author} {\bibfnamefont
  {R.}~\bibnamefont {Moessner}}, \bibinfo {author} {\bibfnamefont {S.~S.}\
  \bibnamefont {Ray}}, \ and\ \bibinfo {author} {\bibfnamefont
  {S.}~\bibnamefont {Bhattacharjee}},\ }\bibfield  {title} {\enquote {\bibinfo
  {title} {{Light-Cone Spreading of Perturbations and the Butterfly Effect in a
  Classical Spin Chain}},}\ }\href {\doibase 10.1103/PhysRevLett.121.024101}
  {\bibfield  {journal} {\bibinfo  {journal} {Phys. Rev. Lett.}\ }\textbf
  {\bibinfo {volume} {121}},\ \bibinfo {pages} {024101} (\bibinfo {year}
  {2018})}\BibitemShut {NoStop}%
\bibitem [{\citenamefont {Sachdev}\ and\ \citenamefont
  {Ye}(1993)}]{PhysRevLett.70.3339}%
  \BibitemOpen
  \bibfield  {author} {\bibinfo {author} {\bibfnamefont {S.}~\bibnamefont
  {Sachdev}}\ and\ \bibinfo {author} {\bibfnamefont {J.}~\bibnamefont {Ye}},\
  }\bibfield  {title} {\enquote {\bibinfo {title} {{Gapless spin-fluid ground
  state in a random quantum Heisenberg magnet}},}\ }\href {\doibase
  10.1103/PhysRevLett.70.3339} {\bibfield  {journal} {\bibinfo  {journal}
  {Phys. Rev. Lett.}\ }\textbf {\bibinfo {volume} {70}},\ \bibinfo {pages}
  {3339--3342} (\bibinfo {year} {1993})}\BibitemShut {NoStop}%
\bibitem [{kit()}]{kitaev}%
  \BibitemOpen
  \href@noop {} {}\bibinfo {note} {A. Y. Kitaev, KITP Program: Entanglement in
  Strongly- Correlated Quantum Matter (2015).}\BibitemShut {Stop}%
\bibitem [{\citenamefont {Banerjee}\ and\ \citenamefont
  {Altman}(2017)}]{PhysRevB.95.134302}%
  \BibitemOpen
  \bibfield  {author} {\bibinfo {author} {\bibfnamefont {S.}~\bibnamefont
  {Banerjee}}\ and\ \bibinfo {author} {\bibfnamefont {E.}~\bibnamefont
  {Altman}},\ }\bibfield  {title} {\enquote {\bibinfo {title} {{Solvable model
  for a dynamical quantum phase transition from fast to slow scrambling}},}\
  }\href {\doibase 10.1103/PhysRevB.95.134302} {\bibfield  {journal} {\bibinfo
  {journal} {Phys. Rev. B}\ }\textbf {\bibinfo {volume} {95}},\ \bibinfo
  {pages} {134302} (\bibinfo {year} {2017})}\BibitemShut {NoStop}%
\bibitem [{\citenamefont {Song}\ \emph {et~al.}(2017)\citenamefont {Song},
  \citenamefont {Jian},\ and\ \citenamefont
  {Balents}}]{PhysRevLett.119.216601}%
  \BibitemOpen
  \bibfield  {author} {\bibinfo {author} {\bibfnamefont {X.-Y.}\ \bibnamefont
  {Song}}, \bibinfo {author} {\bibfnamefont {C.-M.}\ \bibnamefont {Jian}}, \
  and\ \bibinfo {author} {\bibfnamefont {L.}~\bibnamefont {Balents}},\
  }\bibfield  {title} {\enquote {\bibinfo {title} {{Strongly Correlated Metal
  Built from Sachdev-Ye-Kitaev Models}},}\ }\href {\doibase
  10.1103/PhysRevLett.119.216601} {\bibfield  {journal} {\bibinfo  {journal}
  {Phys. Rev. Lett.}\ }\textbf {\bibinfo {volume} {119}},\ \bibinfo {pages}
  {216601} (\bibinfo {year} {2017})}\BibitemShut {NoStop}%
\bibitem [{\citenamefont {Patel}\ and\ \citenamefont
  {Sachdev}(2017)}]{Patel1844}%
  \BibitemOpen
  \bibfield  {author} {\bibinfo {author} {\bibfnamefont {A.~A.}\ \bibnamefont
  {Patel}}\ and\ \bibinfo {author} {\bibfnamefont {S.}~\bibnamefont
  {Sachdev}},\ }\bibfield  {title} {\enquote {\bibinfo {title} {{Quantum chaos
  on a critical Fermi surface}},}\ }\href {\doibase 10.1073/pnas.1618185114}
  {\bibfield  {journal} {\bibinfo  {journal} {Proceedings of the National
  Academy of Sciences}\ }\textbf {\bibinfo {volume} {114}},\ \bibinfo {pages}
  {1844--1849} (\bibinfo {year} {2017})},\ \Eprint
  {http://arxiv.org/abs/http://www.pnas.org/content/114/8/1844.full.pdf}
  {http://www.pnas.org/content/114/8/1844.full.pdf} \BibitemShut {NoStop}%
\bibitem [{\citenamefont {{Kurchan}}(2016)}]{Kurchan2016}%
  \BibitemOpen
  \bibfield  {author} {\bibinfo {author} {\bibfnamefont {J.}~\bibnamefont
  {{Kurchan}}},\ }\bibfield  {title} {\enquote {\bibinfo {title} {{Quantum
  bound to chaos and the semiclassical limit}},}\ }\href@noop {} {\bibfield
  {journal} {\bibinfo  {journal} {ArXiv e-prints}\ } (\bibinfo {year}
  {2016})},\ \Eprint {http://arxiv.org/abs/1612.01278} {arXiv:1612.01278
  [cond-mat.stat-mech]} \BibitemShut {NoStop}%
\bibitem [{\citenamefont {{Scaffidi}}\ and\ \citenamefont
  {{Altman}}(2017)}]{Scaffidi2017}%
  \BibitemOpen
  \bibfield  {author} {\bibinfo {author} {\bibfnamefont {T.}~\bibnamefont
  {{Scaffidi}}}\ and\ \bibinfo {author} {\bibfnamefont {E.}~\bibnamefont
  {{Altman}}},\ }\bibfield  {title} {\enquote {\bibinfo {title} {{Semiclassical
  Theory of Many-Body Quantum Chaos and its Bound}},}\ }\href@noop {}
  {\bibfield  {journal} {\bibinfo  {journal} {ArXiv e-prints}\ } (\bibinfo
  {year} {2017})},\ \Eprint {http://arxiv.org/abs/1711.04768}
  {arXiv:1711.04768} \BibitemShut {NoStop}%
\bibitem [{\citenamefont {Gu}\ \emph {et~al.}(2017)\citenamefont {Gu},
  \citenamefont {Lucas},\ and\ \citenamefont {Qi}}]{Gu2017}%
  \BibitemOpen
  \bibfield  {author} {\bibinfo {author} {\bibfnamefont {Y.}~\bibnamefont
  {Gu}}, \bibinfo {author} {\bibfnamefont {A.}~\bibnamefont {Lucas}}, \ and\
  \bibinfo {author} {\bibfnamefont {X.-L.}\ \bibnamefont {Qi}},\ }\bibfield
  {title} {\enquote {\bibinfo {title} {{Energy diffusion and the butterfly
  effect in inhomogeneous Sachdev-Ye-Kitaev chains}},}\ }\href {\doibase
  10.21468/SciPostPhys.2.3.018} {\bibfield  {journal} {\bibinfo  {journal}
  {SciPost Phys.}\ }\textbf {\bibinfo {volume} {2}},\ \bibinfo {pages} {018}
  (\bibinfo {year} {2017})}\BibitemShut {NoStop}%
\bibitem [{\citenamefont {Blake}(2016)}]{Blake2016}%
  \BibitemOpen
  \bibfield  {author} {\bibinfo {author} {\bibfnamefont {M.}~\bibnamefont
  {Blake}},\ }\bibfield  {title} {\enquote {\bibinfo {title} {{Universal Charge
  Diffusion and the Butterfly Effect in Holographic Theories}},}\ }\href
  {\doibase 10.1103/PhysRevLett.117.091601} {\bibfield  {journal} {\bibinfo
  {journal} {Phys. Rev. Lett.}\ }\textbf {\bibinfo {volume} {117}},\ \bibinfo
  {pages} {091601} (\bibinfo {year} {2016})}\BibitemShut {NoStop}%
\bibitem [{\citenamefont {Blake}\ \emph {et~al.}(2017)\citenamefont {Blake},
  \citenamefont {Davison},\ and\ \citenamefont {Sachdev}}]{Blake2017}%
  \BibitemOpen
  \bibfield  {author} {\bibinfo {author} {\bibfnamefont {M.}~\bibnamefont
  {Blake}}, \bibinfo {author} {\bibfnamefont {R.~A.}\ \bibnamefont {Davison}},
  \ and\ \bibinfo {author} {\bibfnamefont {S.}~\bibnamefont {Sachdev}},\
  }\bibfield  {title} {\enquote {\bibinfo {title} {{Thermal diffusivity and
  chaos in metals without quasiparticles}},}\ }\href {\doibase
  10.1103/PhysRevD.96.106008} {\bibfield  {journal} {\bibinfo  {journal} {Phys.
  Rev. D}\ }\textbf {\bibinfo {volume} {96}},\ \bibinfo {pages} {106008}
  (\bibinfo {year} {2017})}\BibitemShut {NoStop}%
\bibitem [{\citenamefont {{Werman}}\ \emph {et~al.}(2017)\citenamefont
  {{Werman}}, \citenamefont {{Kivelson}},\ and\ \citenamefont
  {{Berg}}}]{Werman2017}%
  \BibitemOpen
  \bibfield  {author} {\bibinfo {author} {\bibfnamefont {Y.}~\bibnamefont
  {{Werman}}}, \bibinfo {author} {\bibfnamefont {S.~A.}\ \bibnamefont
  {{Kivelson}}}, \ and\ \bibinfo {author} {\bibfnamefont {E.}~\bibnamefont
  {{Berg}}},\ }\bibfield  {title} {\enquote {\bibinfo {title} {{Quantum chaos
  in an electron-phonon bad metal}},}\ }\href@noop {} {\bibfield  {journal}
  {\bibinfo  {journal} {ArXiv e-prints}\ } (\bibinfo {year} {2017})},\ \Eprint
  {http://arxiv.org/abs/1705.07895} {arXiv:1705.07895 [cond-mat.str-el]}
  \BibitemShut {NoStop}%
\bibitem [{\citenamefont {{Lucas}}(2017)}]{Lucas2017}%
  \BibitemOpen
  \bibfield  {author} {\bibinfo {author} {\bibfnamefont {A.}~\bibnamefont
  {{Lucas}}},\ }\bibfield  {title} {\enquote {\bibinfo {title} {{Constraints on
  hydrodynamics from many-body quantum chaos}},}\ }\href@noop {} {\bibfield
  {journal} {\bibinfo  {journal} {ArXiv e-prints}\ } (\bibinfo {year}
  {2017})},\ \Eprint {http://arxiv.org/abs/1710.01005} {arXiv:1710.01005
  [hep-th]} \BibitemShut {NoStop}%
\bibitem [{\citenamefont {Balents}\ \emph {et~al.}(2002)\citenamefont
  {Balents}, \citenamefont {Fisher},\ and\ \citenamefont
  {Girvin}}]{PhysRevB.65.224412}%
  \BibitemOpen
  \bibfield  {author} {\bibinfo {author} {\bibfnamefont {L.}~\bibnamefont
  {Balents}}, \bibinfo {author} {\bibfnamefont {M.~P.~A.}\ \bibnamefont
  {Fisher}}, \ and\ \bibinfo {author} {\bibfnamefont {S.~M.}\ \bibnamefont
  {Girvin}},\ }\bibfield  {title} {\enquote {\bibinfo {title}
  {{Fractionalization in an easy-axis Kagome antiferromagnet}},}\ }\href
  {\doibase 10.1103/PhysRevB.65.224412} {\bibfield  {journal} {\bibinfo
  {journal} {Phys. Rev. B}\ }\textbf {\bibinfo {volume} {65}},\ \bibinfo
  {pages} {224412} (\bibinfo {year} {2002})}\BibitemShut {NoStop}%
\bibitem [{\citenamefont {Rehn}\ \emph {et~al.}(2017)\citenamefont {Rehn},
  \citenamefont {Sen},\ and\ \citenamefont {Moessner}}]{Rehn2017}%
  \BibitemOpen
  \bibfield  {author} {\bibinfo {author} {\bibfnamefont {J.}~\bibnamefont
  {Rehn}}, \bibinfo {author} {\bibfnamefont {A.}~\bibnamefont {Sen}}, \ and\
  \bibinfo {author} {\bibfnamefont {R.}~\bibnamefont {Moessner}},\ }\bibfield
  {title} {\enquote {\bibinfo {title} {{Fractionalized {$\mathbb{Z}_{2}$}
  Classical Heisenberg Spin Liquids}},}\ }\href {\doibase
  10.1103/PhysRevLett.118.047201} {\bibfield  {journal} {\bibinfo  {journal}
  {Phys. Rev. Lett.}\ }\textbf {\bibinfo {volume} {118}},\ \bibinfo {pages}
  {047201} (\bibinfo {year} {2017})}\BibitemShut {NoStop}%
\bibitem [{sup()}]{supplemental}%
  \BibitemOpen
  \href@noop {} {}\bibinfo {note} {See Supplemental Material at [URL will be
  inserted by publisher] for the large-N analytics, including Refs. 66-69,
  additional details on the fitting procedure, characteristic raw data fits,
  and variations over initial states.}\BibitemShut {Stop}%
\bibitem [{\citenamefont {Kaneko}(1986)}]{Kaneko1986}%
  \BibitemOpen
  \bibfield  {author} {\bibinfo {author} {\bibfnamefont {K.}~\bibnamefont
  {Kaneko}},\ }\bibfield  {title} {\enquote {\bibinfo {title} {{Lyapunov
  analysis and information flow in coupled map lattices}},}\ }\href {\doibase
  10.1016/0167-2789(86)90149-1} {\bibfield  {journal} {\bibinfo  {journal}
  {Physica D: Nonlinear Phenomena}\ }\textbf {\bibinfo {volume} {23}},\
  \bibinfo {pages} {436--447} (\bibinfo {year} {1986})}\BibitemShut {NoStop}%
\bibitem [{\citenamefont {Deissler}(1984)}]{Deissler1984}%
  \BibitemOpen
  \bibfield  {author} {\bibinfo {author} {\bibfnamefont {R.~J.}\ \bibnamefont
  {Deissler}},\ }\bibfield  {title} {\enquote {\bibinfo {title}
  {{One-dimensional strings, random fluctuations, and complex chaotic
  structures}},}\ }\href {\doibase 10.1016/0375-9601(84)90823-5} {\bibfield
  {journal} {\bibinfo  {journal} {Physics Letters A}\ }\textbf {\bibinfo
  {volume} {100}},\ \bibinfo {pages} {451--454} (\bibinfo {year}
  {1984})}\BibitemShut {NoStop}%
\bibitem [{\citenamefont {Deissler}\ and\ \citenamefont
  {Kaneko}(1987)}]{Deissler1987}%
  \BibitemOpen
  \bibfield  {author} {\bibinfo {author} {\bibfnamefont {R.~J.}\ \bibnamefont
  {Deissler}}\ and\ \bibinfo {author} {\bibfnamefont {K.}~\bibnamefont
  {Kaneko}},\ }\bibfield  {title} {\enquote {\bibinfo {title}
  {{Velocity-dependent Lyapunov exponents as a measure of chaos for open-flow
  systems}},}\ }\href {\doibase 10.1016/0375-9601(87)90581-0} {\bibfield
  {journal} {\bibinfo  {journal} {Physics Letters A}\ }\textbf {\bibinfo
  {volume} {119}},\ \bibinfo {pages} {397--402} (\bibinfo {year}
  {1987})}\BibitemShut {NoStop}%
\bibitem [{\citenamefont {{Khemani}}\ \emph {et~al.}(2018)\citenamefont
  {{Khemani}}, \citenamefont {{Huse}},\ and\ \citenamefont
  {{Nahum}}}]{Khemani2018}%
  \BibitemOpen
  \bibfield  {author} {\bibinfo {author} {\bibfnamefont {V.}~\bibnamefont
  {{Khemani}}}, \bibinfo {author} {\bibfnamefont {D.~A.}\ \bibnamefont
  {{Huse}}}, \ and\ \bibinfo {author} {\bibfnamefont {A.}~\bibnamefont
  {{Nahum}}},\ }\bibfield  {title} {\enquote {\bibinfo {title}
  {{Velocity-dependent Lyapunov exponents in many-body quantum, semi-classical,
  and classical chaos}},}\ }\href@noop {} {\bibfield  {journal} {\bibinfo
  {journal} {ArXiv e-prints}\ } (\bibinfo {year} {2018})},\ \Eprint
  {http://arxiv.org/abs/1803.05902} {arXiv:1803.05902 [cond-mat.stat-mech]}
  \BibitemShut {NoStop}%
\bibitem [{\citenamefont {{Xu}}\ and\ \citenamefont
  {{Swingle}}(2018)}]{Xu2018}%
  \BibitemOpen
  \bibfield  {author} {\bibinfo {author} {\bibfnamefont {S.}~\bibnamefont
  {{Xu}}}\ and\ \bibinfo {author} {\bibfnamefont {B.}~\bibnamefont
  {{Swingle}}},\ }\bibfield  {title} {\enquote {\bibinfo {title} {{Accessing
  scrambling using matrix product operators}},}\ }\href@noop {} {\bibfield
  {journal} {\bibinfo  {journal} {ArXiv e-prints}\ } (\bibinfo {year}
  {2018})},\ \Eprint {http://arxiv.org/abs/1802.00801} {arXiv:1802.00801
  [quant-ph]} \BibitemShut {NoStop}%
\bibitem [{\citenamefont {{Lin}}\ and\ \citenamefont
  {{Motrunich}}(2018)}]{Lin2018a}%
  \BibitemOpen
  \bibfield  {author} {\bibinfo {author} {\bibfnamefont {C.-J.}\ \bibnamefont
  {{Lin}}}\ and\ \bibinfo {author} {\bibfnamefont {O.~I.}\ \bibnamefont
  {{Motrunich}}},\ }\bibfield  {title} {\enquote {\bibinfo {title}
  {{Out-of-time-ordered correlators in short-range and long-range hard-core
  boson models and Luttinger liquid model}},}\ }\href@noop {} {\bibfield
  {journal} {\bibinfo  {journal} {ArXiv e-prints}\ } (\bibinfo {year}
  {2018})},\ \Eprint {http://arxiv.org/abs/1807.08826} {arXiv:1807.08826
  [cond-mat.str-el]} \BibitemShut {NoStop}%
\bibitem [{\citenamefont {Lieb}\ and\ \citenamefont
  {Robinson}(1972)}]{Lieb1972}%
  \BibitemOpen
  \bibfield  {author} {\bibinfo {author} {\bibfnamefont {E.~H.}\ \bibnamefont
  {Lieb}}\ and\ \bibinfo {author} {\bibfnamefont {D.~W.}\ \bibnamefont
  {Robinson}},\ }\bibfield  {title} {\enquote {\bibinfo {title} {{The finite
  group velocity of quantum spin systems}},}\ }\href {\doibase
  10.1007/bf01645779} {\bibfield  {journal} {\bibinfo  {journal}
  {Communications in Mathematical Physics}\ }\textbf {\bibinfo {volume} {28}},\
  \bibinfo {pages} {251--257} (\bibinfo {year} {1972})}\BibitemShut {NoStop}%
\bibitem [{\citenamefont {Marchioro}\ \emph {et~al.}(1978)\citenamefont
  {Marchioro}, \citenamefont {Pellegrinotti}, \citenamefont {Pulvirenti},\ and\
  \citenamefont {Triolo}}]{Marchioro1978}%
  \BibitemOpen
  \bibfield  {author} {\bibinfo {author} {\bibfnamefont {C.}~\bibnamefont
  {Marchioro}}, \bibinfo {author} {\bibfnamefont {A.}~\bibnamefont
  {Pellegrinotti}}, \bibinfo {author} {\bibfnamefont {M.}~\bibnamefont
  {Pulvirenti}}, \ and\ \bibinfo {author} {\bibfnamefont {L.}~\bibnamefont
  {Triolo}},\ }\bibfield  {title} {\enquote {\bibinfo {title} {{Velocity of a
  perturbation in infinite lattice systems}},}\ }\href {\doibase
  10.1007/bf01011695} {\bibfield  {journal} {\bibinfo  {journal} {Journal of
  Statistical Physics}\ }\textbf {\bibinfo {volume} {19}},\ \bibinfo {pages}
  {499--510} (\bibinfo {year} {1978})}\BibitemShut {NoStop}%
\bibitem [{\citenamefont {M\'etivier}\ \emph {et~al.}(2014)\citenamefont
  {M\'etivier}, \citenamefont {Bachelard},\ and\ \citenamefont
  {Kastner}}]{Metivier2014}%
  \BibitemOpen
  \bibfield  {author} {\bibinfo {author} {\bibfnamefont {D.}~\bibnamefont
  {M\'etivier}}, \bibinfo {author} {\bibfnamefont {R.}~\bibnamefont
  {Bachelard}}, \ and\ \bibinfo {author} {\bibfnamefont {M.}~\bibnamefont
  {Kastner}},\ }\bibfield  {title} {\enquote {\bibinfo {title} {{Spreading of
  Perturbations in Long-Range Interacting Classical Lattice Models}},}\ }\href
  {\doibase 10.1103/PhysRevLett.112.210601} {\bibfield  {journal} {\bibinfo
  {journal} {Phys. Rev. Lett.}\ }\textbf {\bibinfo {volume} {112}},\ \bibinfo
  {pages} {210601} (\bibinfo {year} {2014})}\BibitemShut {NoStop}%
\bibitem [{\citenamefont {Roberts}\ and\ \citenamefont
  {Swingle}(2016)}]{Roberts2016}%
  \BibitemOpen
  \bibfield  {author} {\bibinfo {author} {\bibfnamefont {D.~A.}\ \bibnamefont
  {Roberts}}\ and\ \bibinfo {author} {\bibfnamefont {B.}~\bibnamefont
  {Swingle}},\ }\bibfield  {title} {\enquote {\bibinfo {title} {{Lieb-Robinson
  Bound and the Butterfly Effect in Quantum Field Theories}},}\ }\href
  {\doibase 10.1103/PhysRevLett.117.091602} {\bibfield  {journal} {\bibinfo
  {journal} {Phys. Rev. Lett.}\ }\textbf {\bibinfo {volume} {117}},\ \bibinfo
  {pages} {091602} (\bibinfo {year} {2016})}\BibitemShut {NoStop}%
\bibitem [{\citenamefont {Garanin}\ and\ \citenamefont
  {Canals}(1999)}]{Garanin1999}%
  \BibitemOpen
  \bibfield  {author} {\bibinfo {author} {\bibfnamefont {D.~A.}\ \bibnamefont
  {Garanin}}\ and\ \bibinfo {author} {\bibfnamefont {B.}~\bibnamefont
  {Canals}},\ }\bibfield  {title} {\enquote {\bibinfo {title} {{Classical Spin
  Liquid: Exact Solution for the Infinite-component Antiferromagnetic Model on
  the Kagom\'e Lattice}},}\ }\href {\doibase 10.1103/PhysRevB.59.443}
  {\bibfield  {journal} {\bibinfo  {journal} {Phys. Rev. B}\ }\textbf {\bibinfo
  {volume} {59}},\ \bibinfo {pages} {443--456} (\bibinfo {year}
  {1999})}\BibitemShut {NoStop}%
\bibitem [{\citenamefont {Isakov}\ \emph {et~al.}(2004)\citenamefont {Isakov},
  \citenamefont {Gregor}, \citenamefont {Moessner},\ and\ \citenamefont
  {Sondhi}}]{Isakov2004}%
  \BibitemOpen
  \bibfield  {author} {\bibinfo {author} {\bibfnamefont {S.~V.}\ \bibnamefont
  {Isakov}}, \bibinfo {author} {\bibfnamefont {K.}~\bibnamefont {Gregor}},
  \bibinfo {author} {\bibfnamefont {R.}~\bibnamefont {Moessner}}, \ and\
  \bibinfo {author} {\bibfnamefont {S.~L.}\ \bibnamefont {Sondhi}},\ }\bibfield
   {title} {\enquote {\bibinfo {title} {{Dipolar Spin Correlations in Classical
  Pyrochlore Magnets}},}\ }\href {\doibase 10.1103/PhysRevLett.93.167204}
  {\bibfield  {journal} {\bibinfo  {journal} {Phys. Rev. Lett.}\ }\textbf
  {\bibinfo {volume} {93}},\ \bibinfo {pages} {167204} (\bibinfo {year}
  {2004})}\BibitemShut {NoStop}%
\bibitem [{\citenamefont {Taillefumier}\ \emph {et~al.}(2014)\citenamefont
  {Taillefumier}, \citenamefont {Robert}, \citenamefont {Henley}, \citenamefont
  {Moessner},\ and\ \citenamefont {Canals}}]{Taillefumier2014}%
  \BibitemOpen
  \bibfield  {author} {\bibinfo {author} {\bibfnamefont {M.}~\bibnamefont
  {Taillefumier}}, \bibinfo {author} {\bibfnamefont {J.}~\bibnamefont
  {Robert}}, \bibinfo {author} {\bibfnamefont {C.~L.}\ \bibnamefont {Henley}},
  \bibinfo {author} {\bibfnamefont {R.}~\bibnamefont {Moessner}}, \ and\
  \bibinfo {author} {\bibfnamefont {B.}~\bibnamefont {Canals}},\ }\bibfield
  {title} {\enquote {\bibinfo {title} {{Semiclassical Spin Dynamics of the
  Antiferromagnetic Heisenberg Model on the Kagome Lattice}},}\ }\href
  {\doibase 10.1103/PhysRevB.90.064419} {\bibfield  {journal} {\bibinfo
  {journal} {Phys. Rev. B}\ }\textbf {\bibinfo {volume} {90}},\ \bibinfo
  {pages} {064419} (\bibinfo {year} {2014})}\BibitemShut {NoStop}%
\bibitem [{\citenamefont {Conlon}\ and\ \citenamefont
  {Chalker}(2009)}]{Conlon2009}%
  \BibitemOpen
  \bibfield  {author} {\bibinfo {author} {\bibfnamefont {P.~H.}\ \bibnamefont
  {Conlon}}\ and\ \bibinfo {author} {\bibfnamefont {J.~T.}\ \bibnamefont
  {Chalker}},\ }\bibfield  {title} {\enquote {\bibinfo {title} {{Spin Dynamics
  in Pyrochlore Heisenberg Antiferromagnets}},}\ }\href {\doibase
  10.1103/PhysRevLett.102.237206} {\bibfield  {journal} {\bibinfo  {journal}
  {Phys. Rev. Lett.}\ }\textbf {\bibinfo {volume} {102}},\ \bibinfo {pages}
  {237206} (\bibinfo {year} {2009})}\BibitemShut {NoStop}%
\end{thebibliography}%
%  \input{butterfly.bbl}

 % %%%%%%%%%% Merge with supplemental materials %%%%%%%%%%
 \cleardoublepage

 \begin{center}
   \textbf{\large Supplemental Material:}
 \end{center}
 % %%%%%%%%%% Merge with supplemental materials %%%%%%%%%%
 % %%%%%%%%%% Prefix a "S" to all equations, figures, tables and reset the
 % %%%%%%%%%% counter %%%%%%%%%%
 \setcounter{equation}{0} \setcounter{figure}{0} \setcounter{table}{0}
 \setcounter{page}{1} \makeatletter
 \renewcommand{\theequation}{S\arabic{equation}}
 \renewcommand{\thefigure}{S\arabic{figure}} \renewcommand{\bibnumfmt}[1]{[S#1]}
 % \renewcommand{\citenumfont}[1]{S#1}
 % %%%%%%%%%% Prefix a "S" to all equations, figures, tables and reset the
 % %%%%%%%%%% counter %%%%%%%%%%

 \section{Dynamics of the $\mathbb{Z}_2$ spin liquid}

 \subsection{Large-N analytics}
 The large-N limit relaxes the condition of unit length $O(N)$ spins in the
 limit of $N \rightarrow \infty$ \cite{Garanin1999}. It reproduces well the
 static properties of the highly frustrated kagome and pyrochlore Heisenberg
 models \cite{Isakov2004,Taillefumier2014}. Its extension via a stochastic
 Langevin dynamics allows accurate predictions also for the dynamics of the
 classical pyrochlore and kagome models \cite{Conlon2009,Taillefumier2014}.

 Eventhough the quantitative agreement has been shown to be slightly worse for
 the $\mathbb{Z}_2$ model under consideration here \cite{Rehn2017}, it provides
 an analytically tractable starting point from which to approach the full model
 dynamics.

 \sectionn{Large-N calculation} In the large-N calculation the soft spins follow
 the (unnormalised) probability distribution $e^{-\beta E}$ with the energy
 \begin{equation}
   \beta E = \frac{1}{2} \sum_i \lambda s_i^2 + \frac{1}{2} \beta J \sum_{\alpha} l_{\alpha}^2
 \end{equation}
 where $l_{\alpha}=\sum_{i \in \alpha} s_i$ is the sum of the ``soft'' spins
 $s_i$ over the hexagon $\alpha$ and $\lambda$ is a lagrange multiplier ensuring
 the length constraint $\left< s_i^2\right> = 1/3$ (Heisenberg spins).

 We may rewrite the interaction term as $\sum_{i,j} s_i \left( A_{ij}+2
   \delta_{ij} \right) s_j$ where $A_{ij}$ is the connectivity matrix of the
 model. We call $M=\left( A_{ij} +2 \delta_{ij} \right)$ the interaction matrix.

 Since the model is translationally invariant, the eigenbasis is labelled by a
 momentum $\vect{q}$ and a sublattice index $\nu \in \{1,2,3\}$. We obtain two
 flat bands $\nu_{1,2}(\vect{q})=0$ and one dispersive gapped band
 $\nu_3(\vec{q})$.

 Dynamics is introduced via the Langevin equation
 \begin{equation}
   \frac{d s_i(t)}{dt} = \Gamma \sum_l \left( A_{il} - z \delta_{il} \right) \frac{\partial E}{\partial s_l}  + \zeta_i(t)
 \end{equation}
 with the coordination number $z$ ($z=10$ in this model), a noise term
 $\zeta_i(t)$ and a free constant $\Gamma$ determining the overall timescale of
 dynamical processes.

 Solving this equation in the eigenbasis $\tilde{s}(\vect{q})$ of the
 interaction matrix $M$ we obtain
 \begin{equation}
   \left<\tilde{s}^{\mu}_{\vect{q}}(t) \tilde{s}^{\nu}_{-\vect{q}}(0) \right> = \frac{\delta_{\mu,\nu}T}{J v_{\mu}+\lambda T} e^{- \Gamma \left( 12 -v_{\mu} \right) \left(J v_{\mu} +\lambda T \right) t}
 \end{equation}

 with a characteristic decay rate $\kappa_{\mu} = \Gamma \left( 12 -v_{\mu}
 \right) \left(J v_{\mu} +\lambda T \right)$.

 The dynamical structure factor is given by
 \begin{align}
   S(\vect{q},t)&=\sum_{\nu \mu}\left< s^\mu _{\vect{q}}(t) s^{\nu}_{-\vect{q}}(0)\right> \\
                &= \sum_{\alpha} \left( \sum_{\mu \nu} U^{\mu \alpha}_{\vect{q}} U^{\nu \alpha}_{-\vect{q}}\right) \left<\tilde{s}^{\alpha}_{\vect{q}}(t) \tilde{s}^{\alpha}_{-\vect{q}}(0) \right> \\
                &= \sum_{\alpha} g^{\alpha}_{\vect{q}} \left<\tilde{s}^{\alpha}_{\vect{q}}(t) \tilde{s}^{\alpha}_{-\vect{q}}(0) \right> 
 \end{align}
 with the factors $g^{\alpha}_{\vect{q}}$ defined via the matrix of eigenvectors
 $U^{\mu \nu}_{\vect{q}}$. These factors satisfy the sum rule $\sum_{\alpha}
 g^{\alpha} = 3$.

 \sectionn{Autocorrelation} The autocorrelation function shows exponential decay
 $A(t) = e^{-\kappa t}$ in the low-temperature limit with $\kappa = 12 \Gamma
 \lambda T $, i.e. a linear dependence on temperature.

 \sectionn{Structure factor} In the low-temperature limit the dynamical
 structure factor at generic wave-vector is exponentially decaying with a
 $\vect{q}$-independent decay rate scaling with $T$, $\kappa(\vect{q}) = 12
 \Gamma \lambda T$.

 Around $\vect{q}=0$ the full weight is in the dispersive band, and we can
 extract a diffusion constant from $\kappa(\vect{q}) = D q^2$ with $D= 9 \,
 \Gamma J$, i.e. a temperature independent diffusion constant.

 \section{Numerics}

\subsection{Structure factor}
We consider the dynamical structure factor $S(\vect{q},t) = \sum_{ij}
e^{i\vect{q}\cdot(\vect{r}_i - \vect{r}_j)} \langle \vect{S}_i(t) \cdot
\vect{S}_j(0) \rangle$ and its fourier transform $S(\vect{q},\omega)$ which
provides spatially and frequency resolved information on the dynamics.

\begin{figure}
  \begin{minipage}{0.99\columnwidth}
    \includegraphics[width=.99\columnwidth]{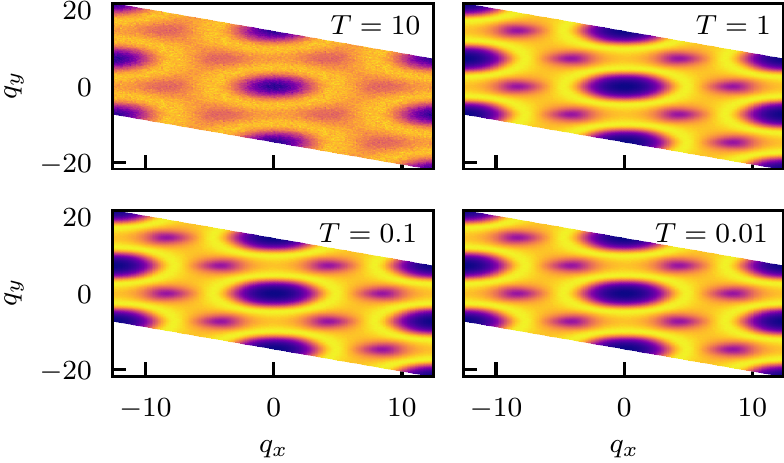}
  \end{minipage}
  \caption{Static structure factor $S(\mathbf{q},t=0)$ for temperatures $T=10$
    down to $T=0.01$ as indicated in the figure. 
    \label{fig:supp_static_structure}}
\end{figure}
\sectionn{Static structure factor} The static structure factor $S(\vect{q},t=0)$
in Fig.~\ref{fig:supp_static_structure} for temperatures $T=0.01,0.1,1,10$ shows
the transition from the high-temperature paramagnet to the spin liquid regime at
low temperatures.

The static structure factor remains essentially unchanged below $T \sim 0.1-1$.
In particular, it shows no pinch points or lines and no sign of ordering. The
results in the spin-liquid regime are in good agreement with the predictions of
the large-N calculations (not shown).

\sectionn{Dynamical structure factor} Since the dynamics, Eq.~\ref{eq:bloch_eq},
conserves the total magnetisation, we expect diffusion to occur at small
wavevectors.

\begin{figure}
  \begin{minipage}{0.99\columnwidth}
    \includegraphics[width=.99\columnwidth]{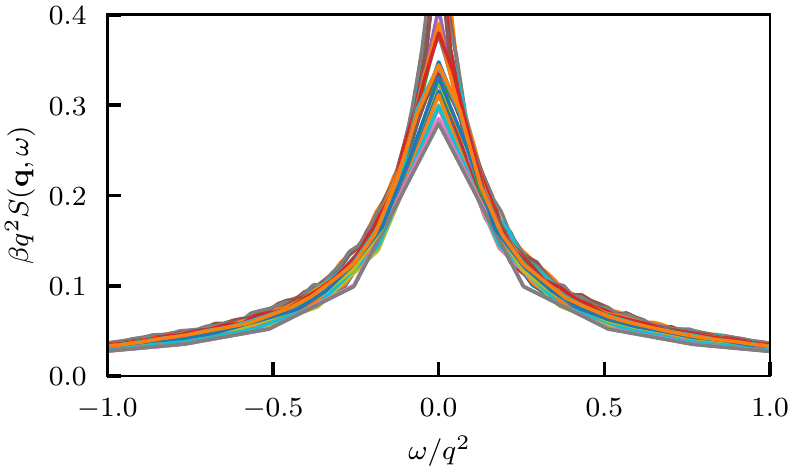}
  \end{minipage}
   \caption{Scaling collapse of the dynamical structure factor
     $S(\vect{q},\omega)$ for $\vect{q}$ around the $\Gamma$-point and
     temperatures $T=0.01,0.02,0.04,0.1$.
     \label{fig:supp_diffusion_scaling_collapse}}
 \end{figure}

To test this expectation we perform a scaling collapse of the dynamical
structure factor via
\begin{equation}
  \beta q^2 S(\vect{q},\omega) \sim  \frac{D}{(\omega/q^2)^2 + D^2}
\end{equation}
appropriate diffusion in 2D. We find an approximate scaling collapse of this
form in Fig.~\ref{fig:supp_diffusion_scaling_collapse}.

 To extract the diffusion constant we fit the dynamic structure factor via
 $S(\vect{q},\omega) \sim 1/(\omega^2 + \kappa(\vect{q})^2)$, corresponding to
 an exponentially decaying dynamical structure factor with decay rate
 $\kappa(\vect{q})$, i.e. $S(\vect{q},t) \sim e^{-\kappa(\vect{q})t}$. For
 diffusive behaviour we expect $\kappa(\vect{q}) = D q^2$ for momenta $q$ close
 to the $\Gamma$-point.

\begin{figure}
   \begin{minipage}{0.99\columnwidth}
     \includegraphics[width=.99\columnwidth]{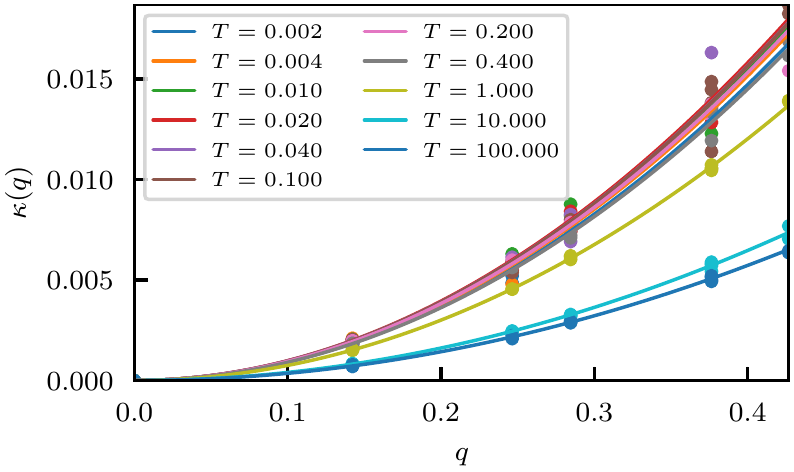}
   \end{minipage}
   \caption{Extraction of the diffusion constant from the dynamical structure.
     Fits to $\kappa(q) = D q^2$ for a range of temperatures $T=0.002$ up to
     $T=100$. The data separates into the high temperature regime $T>1$ and the
     low temperature spin-liquid regime $T<0.1$.
     \label{fig:supp_diffusion_constant_fits}}
 \end{figure}

 Fig.~\ref{fig:supp_diffusion_constant_fits} presents the results for
 $\kappa(q)$ and the quadratic fits to extract the diffusion constant $D$.
 Already on the level of the raw data we observe a clear separation into the
 high-temperature paramagnetic phase $T>1$ and the low-temperature spin-liquid
 regime $T<0.1$.

 We also note that with decreasing temperature the range of validity of the
 quadratic fit shrinks which limits the extraction of the diffusion constant to
 temperatures $T \ge 0.002$ on the available system sizes, and results in
 increasing uncertainties at lower temperatures.
 
 % The obtained diffusion constant $D$ in Fig.~\ref{fig:supp_relaxation}(b)
 % shows saturation to a constant value at temperatures $T<0.1$ once the
 % spin-liquid regime is reached and is not (or only very weakly) temperature
 % dependent below. This is another indicator that there is no ordering
 % transition which would lead to a diverging diffusion constant at the
 % transition point.

 \section{Butterfly effect in the $\mathbb{Z}_2$ spin liquid}

 \subsection{Fitting the full propagating wavefront}
 As stated in the main text the de-correlator $D(x,t)$ is fit well by the
 scaling form
 \begin{equation}
   D(x,t) \sim \exp [2 \mu (1- (v/v_b)^{\nu}) t]
   \label{eq:supp_D_scaling}
 \end{equation}
 with the Lyapunov exponent $\mu$, the butterfly velocity $v_b$ and an exponent
 $\nu$, which will all generically be temperature dependent.

 \begin{figure}
   \begin{minipage}{0.49\columnwidth}
     \includegraphics[width=.99\columnwidth]{{./figures/z2_cross_cor_final_fit_T_100_eps_0.0001}.pdf}
   \end{minipage}
   \begin{minipage}{0.49\columnwidth}
     \includegraphics[width=.99\columnwidth]{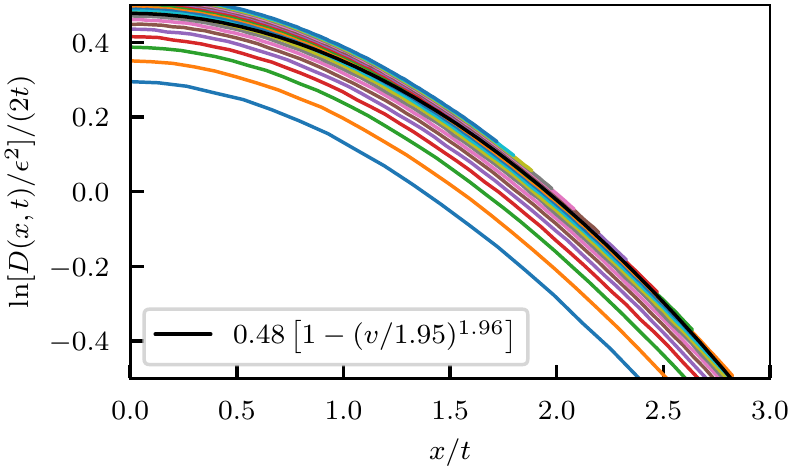}
   \end{minipage}

   \begin{minipage}{0.49\columnwidth}
     \includegraphics[width=.99\columnwidth]{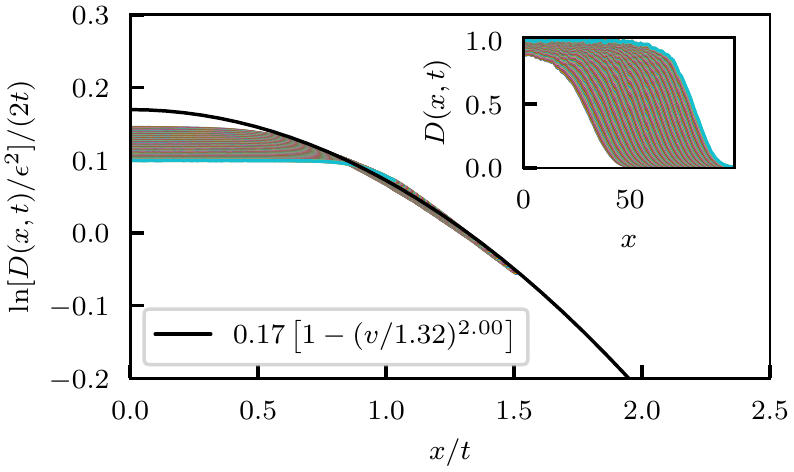}
   \end{minipage}
   \begin{minipage}{0.49\columnwidth}
     \includegraphics[width=.99\columnwidth]{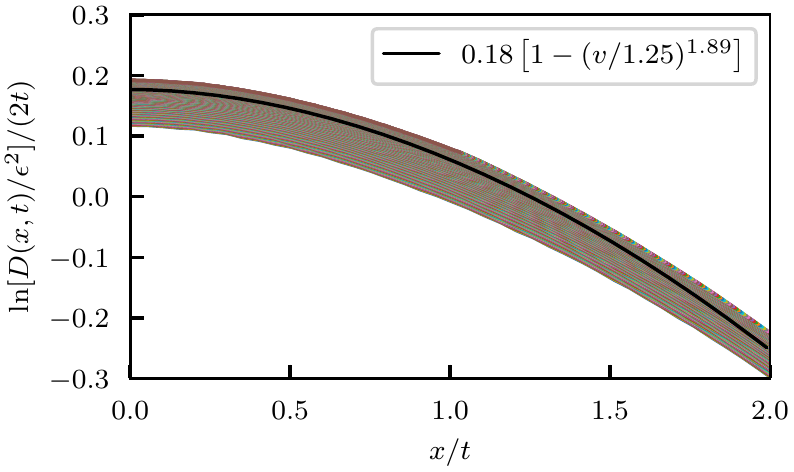}
   \end{minipage}
   \caption{Scaling form and ballistic propagation of wave-fronts in the
    de-correlator $D(x,t)$. Non-linear dynamics with $\epsilon=10^{-4}$
    (left) versus linearised dynamics (right) and $T=100$ (top) versus $T=0.1$ (bottom).
    The data at the wave-front $v =x/t \sim v_b$ well fits Eq.~\ref{eq:supp_D_scaling}
    with the parameters given in the box. The inset shows the unscaled data
    versus distance $x$ at different time-slices $t$ demonstrating the ballistic
    propagation of the wave-front after an initial growth period. 
    \label{fig:supp_D_scaling_form}}
\end{figure}
In Fig.~\ref{fig:supp_D_scaling_form} we show typical fits of the scaling form
to the results of the de-correlator $D(x,t)$ for the non-linear (left row) and
linearised dynamics (right row) for temperatures $T=100$ (top) and $T=0.1$
(bottom).

The non-linear data shows saturation effects inside the light-cone for $v<v_b$
when the decorrelation reaches $D \sim 1$. Thus, it only fits the scaling form
around $v \sim v_b$ and the estimate for $\mu$ is biased to smaller values. The
linearised dynamics shows no saturation effects, and therefore provides
considerably less spread around the scaling form. Moreover, the remaining spread
is further reduced when performing the simulations on larger system sizes.

% The different scaling of $\mu$ and $v_b$ with temperature shows itself in two
% effects here. Firstly, we observe that the wavefront as a function of $x$
% broadens with decreasing temperatures as discussed above.

Smaller temperatures inherently require larger systems for a wave front to build
up since the perturbation propagates relatively faster than it grows, thus,
reaching the boundaries of the system before it saturates at the initial site.
This limits the temperature range we can reliably perform fits to the full
wavefronts to $T >0.02$ on systems up to $L=101$.

% \begin{figure}
%   \begin{minipage}{0.95\columnwidth}
%     \includegraphics[width=.99\columnwidth]{{./figures/z2_cross_cor_butterfly_full_fit_final_T}.pdf}
%   \end{minipage}
%   \caption{Result of fits to  $D(x,t) \sim \exp [2 \mu t (1- (v/v_b)^{\alpha})]$
%     versus temperature for linearised dynamics (squares) and non-linear dynamics
%     with $\epsilon=0.0001$
%     (circles), exponent $\alpha$ (dashed), butterfly speed $v_b$ (solid) and Lyapunov $\mu$ (dashed-dotted).
%     \label{fig:supp_cross_cor_fit_results}}
%   \end{figure}
%   The full results resulting from the fits to the scaling for the butterfly
%   speed $v_b$, the Lyapunov exponent $\mu$ and the exponent $\nu$ versus
%   temperature $T$ are shown in Fig.~\ref{fig:supp_cross_cor_fit_results}

%   We observe that fits to the full wave front generally result in slightly
%   lower estimates of $\mu$ in the non-linear dynamics due to the saturation of
%   the decorrelation at $v=0$ affecting the global fit. Consequently, the
%   estimates for the butterfly velocity $v_b$ are slightly higher.

The linearised dynamics show an extracted exponent $\nu=2$ at $T=100$, which
then decreases slightly to $\nu \sim 1.9$ at $T=0.1$. The non-linear dynamics
are consistent with a constant exponent $\nu = 2$ over the full temperature
range considered, however, can also be fit with the exponent extracted from the
linearised dynamics.

\subsection{Convergence with $\epsilon$}
\begin{figure}
  \begin{minipage}{0.99\columnwidth}
    \includegraphics[width=.99\columnwidth]{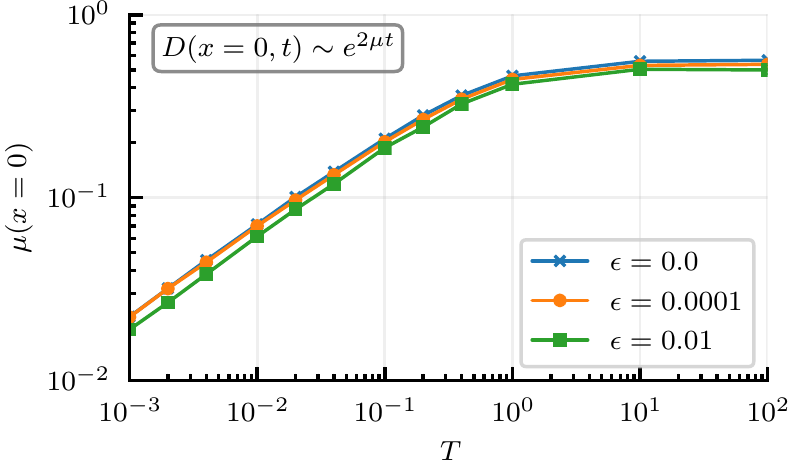}
  \end{minipage}
  \caption{Lyapunov Exponent $\mu$ versus temperature $T$ on a log-log scale, extracted from an exponential fit to the
    de-correlator at $x=0$ via $D(x=0,t) \sim \exp [2 \mu t]$ for the
    non-linear dynamics with $\epsilon=10^{-4},10^{-2}$ and the linearised
    dynamics $\epsilon=0$.
    \label{fig:supp_cross_cor_mu_x0}
  }
\end{figure}
Fitting the behaviour of $D(x=0,t)$ via $D(x=0,t) \sim e^{2 \mu t} $ allows to
extract the (leading) Lyapunov exponent $\mu$. The extracted Lyapunov exponent
is shown in Fig.~\ref{fig:supp_cross_cor_mu_x0} versus temperature. We observe
convergence of the extracted Lyapunov exponent with decreasing $\epsilon$
towards the results of the linearised equations across the whole temperature
range. We note that this is important for two reasons. Firstly, it shows that
our results for $\epsilon=10^{-4}$ are already quite close to the limit of
vanishing perturbation strength. Secondly, it implies that the linearised
dynamics indeed correctly captures the behaviour of the decorrelator also for
finite, but small perturbation strengths.

We may determine the butterfly-speed from the propagation of the wavefront: We
define the arrival time $t_{D_0}(x)$ at distance $x$ at which the de-correlator
$D(x,t)$ exceeds a given threshold $D_0$. For ballistic propagation we expect a
linear relation with $x = v_b \, t_{D_0}$.
 
\begin{figure}
  \begin{minipage}{0.99\columnwidth}
    \includegraphics[width=.99\columnwidth]{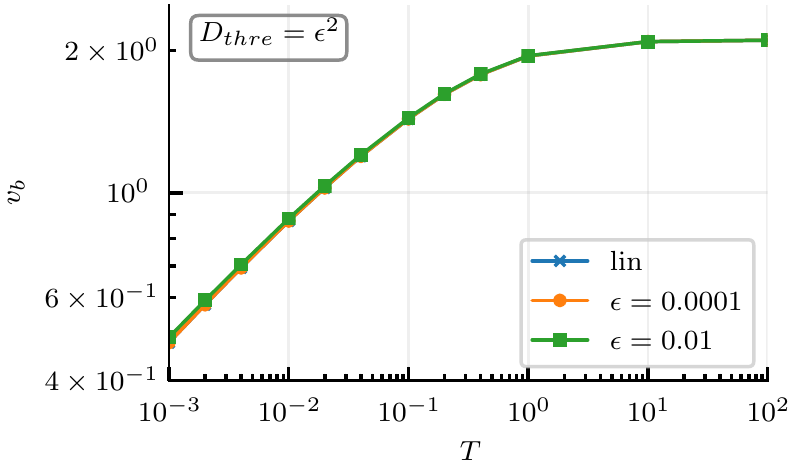}
  \end{minipage}
  \caption{Butterfly speed $v_b$ versus temperature $T$ on a log-log scale, extracted from a linear fit to the arrival times
    $x = v_b t_{D_0}$ defined via $D(x,t_{D_0}) > D_0$.
    \label{fig:supp_cross_cor_vb}}
\end{figure}

For sufficiently small thresholds $D_0 \sim \epsilon^2$ we observe the expected
linear behaviour of the arrival times with distance $t_{D_0} = x/v_b$, at least
for sites $x$ sufficiently removed from the initially perturbed site. Moreover,
choosing $D_0 = \epsilon^2$ we obtain results for the butterfly speed $v_b$
independent of the chosen perturbation strength, for sufficiently small
$\epsilon$, and in agreement with the linearised dynamics as shown in
Fig.~\ref{fig:supp_cross_cor_vb}.

% At large temperatures the velocity saturates at $v_b= 2$. At low temperatures
% after the system enters the spin-liquid regime below $T<0.1$ we observe
% powerlaw behaviour $v_b \sim T^{0.48}$.

\subsection{Variation over initial states}
In this section we consider the dependence of the extracted quantities on the
initial states keeping information for all $10^4$ simulated states, but
restricting to smaller sizes of $L=51$.

The sample-to-sample variation allows us to determine whether the mean
characterises the full state-manifold or whether states at a given temperature
might behave differently.

\begin{figure}
  \begin{minipage}{0.99\columnwidth}
    \includegraphics[width=.99\columnwidth]{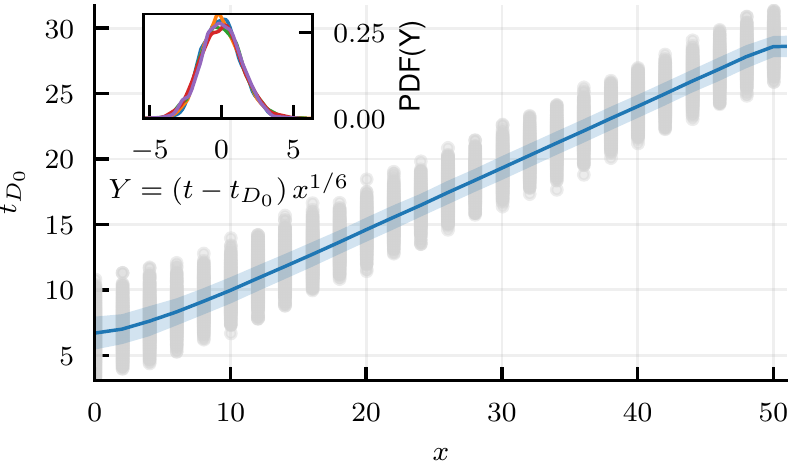}
  \end{minipage}
  \begin{minipage}{0.99\columnwidth}
    \includegraphics[width=.99\columnwidth]{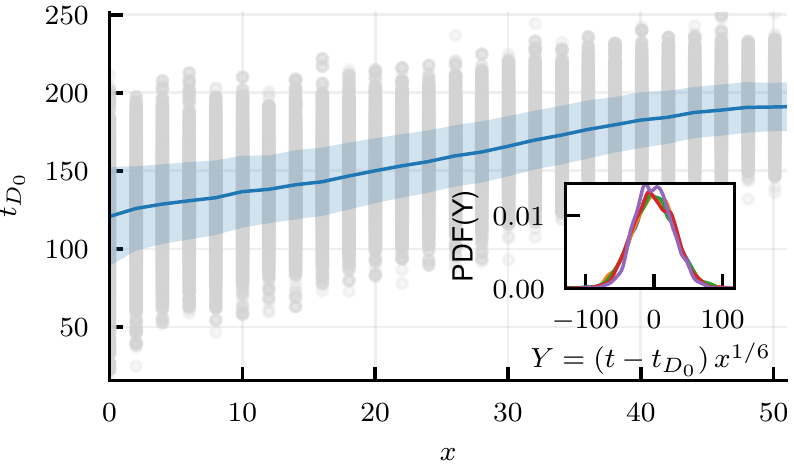}
  \end{minipage}
   \caption{Arrival times $t_{D_0}$ as a function of x for $D_0=10 \epsilon^2$
     and $\epsilon=10^{-4}$ for temperatures $T=100$ (left) and $T=0.001$
     (right). Gray scatter is the variation over different initial
     spin-configuration, solid blue line the mean value, and the blue shading
     the standard deviation of the data. The inset shows the distribution of
     $t_{D_0}$ at $x=5,10,15,20,25$ with an approximate collapse on scaling with
     $x^{1/6}$. Results obtained on a $L=51$ system.
     \label{fig:supp_t_D0_sample}}
 \end{figure}
 We first consider the variation of the arrival times $t_{D_0}$ in
 Fig.~\ref{fig:supp_t_D0_sample}. The relation of $t_{D_0}$ is linear for all
 samples apart from boundary effects, either at the initial perturbed site or at
 half the system size when periodic boundary conditions affect the results.

 We observe some scatter in the arrival times, increasing considerably at lower
 temperature. However, the variance of the arrival times actually appears to
 decrease with increasing distance $x$ from the perturbed site, or equivalently
 with time. This is confirmed in the inset by the scaling collapse of the data
 for $t_{D_0}$ with $x^{1/6}$.

 However, we emphasise that this might only be true for the accessible times,
 and this initial decrease could be a transient effect, after which the
 asymptotic long-time limit could show different behaviour.

 \begin{figure}
   \begin{minipage}{0.99\columnwidth}
     \includegraphics[width=.99\columnwidth]{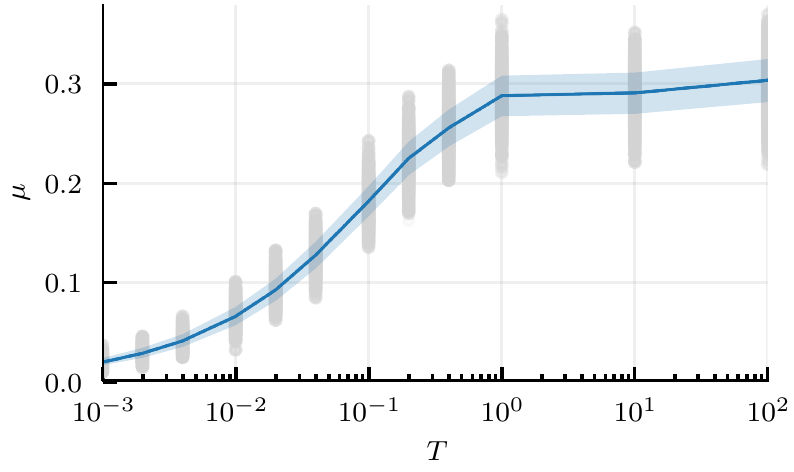}
   \end{minipage}
  \caption{Lyapunov exponent $\mu$ versus temperature $T$, extracted from an exponential fit to the
    de-correlator at $x=0$ via $D(x=0,t) \sim \exp [2 \mu t]$ for different
    initial states. Gray scatter is the variation over initial states, the solid
    line the mean, and the blue shading marks the standard deviation of the
    data. Results obtained for the non-linear dynamics with $\epsilon=10^{-4}$ on a $L=51$ system.
    \label{fig:supp_mu_sample}}
\end{figure}
Next, we consider the extraction of the Lyapunov exponent from the de-correlator
at $x=0$ via $D(x=0,t) \sim \exp [2 \mu t]$. In Fig.~\ref{fig:supp_mu_sample} we
show the variation of $\mu$ at different temperatures $T$ when performing this
fit for different initial states individually instead of on the mean of the
data. The distribution of $\mu$ over initial states is approximately gaussian at
all temperatures and well characterised by its mean and variance, both
decreasing with decreasing temperatures.

We note that the average of the extracted Lyapunov exponents over initial states
differs from the Lyapunov exponent extracted from the averaged data, since the
former is essentially the average of a logarithm, whereas the later is the
logarithm of the average.

% From these results we conclude that we can reliably extract the butterfly
% velocity and the Lyapunov exponent from the mean of the data and that this
% characterises the full thermodynamic ensemble.

\end{document}